\documentclass[journal]{IEEEtran}
\ifCLASSINFOpdf \else \fi

\usepackage{graphicx}

\usepackage[cmex10]{amsmath}
\usepackage{amssymb}

\usepackage{array}
\ifCLASSOPTIONcompsoc
 \usepackage[caption=false,font=normalsize,labelfont=sf,textfont=sf]{subfig}
\else
 \usepackage[caption=false,font=footnotesize]{subfig}
\fi
\usepackage{url}
\usepackage[noadjust]{cite}

\usepackage{cite}
\usepackage{amsmath,amssymb,amsfonts}
\usepackage{graphicx}
\usepackage{textcomp}
\usepackage{listings}
\usepackage{stfloats}
\usepackage{amssymb}
\usepackage{bm}
\usepackage{color}
\usepackage{algorithm}
\usepackage{algpseudocode}
\usepackage{amsmath}
\usepackage{authblk}
\usepackage{booktabs}
\usepackage{hyperref}
\usepackage[table,x11names]{xcolor}
\usepackage{float}
\usepackage{multirow}
\def\BibTeX{{\rm B\kern-.05em{\sc i\kern-.025em b}\kern-.08em
    T\kern-.1667em\lower.7ex\hbox{E}\kern-.125emX}}

\title{Motion-compensated MR CINE reconstruction with reconstruction-driven motion estimation}
\author{Jiazhen Pan, Wenqi Huang, Daniel Rueckert, \IEEEmembership{Fellow, IEEE}, Thomas Küstner, \IEEEmembership{Member, IEEE} and Kerstin Hammernik
\thanks{This work was supported in part by the European Research Council (Grant Agreement no. 884622).}
\thanks{Thomas Küstner and Kerstin Hammernik contributed equally.}
\thanks{Jiazhen Pan, Wenqi Huang, Daniel Rueckert and Kerstin Hammernik are with Klinikum Rechts der Isar, Technical University of Munich, Munich, Germany (e-mail: jiazhen.pan@tum.de; wenqi.huang@tum.de; daniel.rueckert@tum.de; k.hammernik@tum.de).}
\thanks{Thomas Küstner is with the University Hospital of Tuebingen, Department of Diagnostic and Interventional Radiology, Medical Image and Data Analysis (MIDAS.lab),
Tübingen, Germany (e-mail: thomas.kuestner@med.uni-tuebingen.de).}
\thanks{Daniel Rueckert and Kerstin Hammernik are also with the Department of Computing, Imperial College London, London, United Kingdom.}}

\begin{document}

% make the title area
\maketitle

\begin{abstract}
% alternative opening sentence to replace the first sentence
%In cardiac CINE, cardiac motion serves as a key component for solving the reconstruction problem. Motion-compensated MR reconstruction (MCMR) incorporating this motion information between frames can address highly undersampled acquisitions while other methods fail.

In cardiac CINE, motion-compensated MR reconstruction (MCMR) is an effective approach to address highly undersampled acquisitions by incorporating motion information between frames. In this work, we propose a novel perspective for addressing the MCMR problem and a more integrated and efficient solution to the MCMR field. Contrary to state-of-the-art (SOTA) MCMR methods which break the original problem into two sub-optimization problems, i.e. motion estimation and reconstruction, we formulate this problem as a single entity with one single optimization. Our approach is unique in that the motion estimation is directly driven by the ultimate goal, reconstruction, but not by the canonical motion-warping loss (similarity measurement between motion-warped images and target images). We align the objectives of motion estimation and reconstruction, eliminating the drawbacks of artifacts-affected motion estimation and therefore error-propagated reconstruction. Further, we can deliver high-quality reconstruction and realistic motion without applying any regularization/smoothness loss terms, circumventing the non-trivial weighting factor tuning. We evaluate our method on two datasets: 1) an in-house acquired 2D CINE dataset for the retrospective study and 2) the public OCMR cardiac dataset for the prospective study. The conducted experiments indicate that the proposed MCMR framework can deliver artifact-free motion estimation and high-quality MR images even for imaging accelerations up to 20x, outperforming SOTA non-MCMR and MCMR methods in both qualitative and quantitative evaluation across all experiments. The code is available at \textcolor{magenta}{\href{https://github.com/JZPeterPan/MCMR-Recon-Driven-Motion}{https://github.com/JZPeterPan/MCMR-Recon-Driven-Motion}}.

\end{abstract}


\section{Introduction}
\label{sec:introduction}
\begin{figure}
	\centering
	\centerline{\includegraphics[width=\columnwidth]{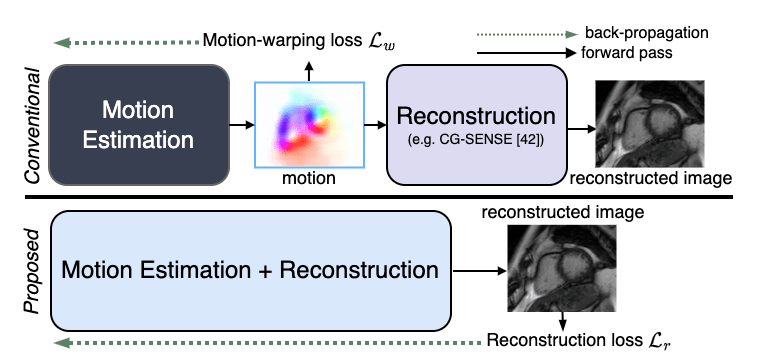}}
	\caption{The difference between the proposed MCMR framework (bottom) and the conventional MCMR work (top) is shown. The conventional approaches divide the original MCMR problem into two sub-optimization problems: motion estimation and reconstruction. Its motion estimation is optimized by minimizing the intermediate motion-warping loss (brightness similarity measurement between motion-warped images and target images) and if deep learning is used, the motion prediction back-propagation is only exerted on the motion estimation part. In contrast, we develop a deep learning-based framework that predicts the motion from the perspective of our ultimate goal: reconstruction. We discard using any intermediate motion-warping loss. The back-propagation is performed through the whole pipeline and reconstruction-driven motion estimation is established.}
	\label{fig_overview}
\end{figure}

\IEEEPARstart{C}{ine} cardiac magnetic resonance imaging (CMR) serves as a versatile tool for characterizing cardiac morphology and assessing cardiac function. Quantitative indicators such as volume and ejection fraction can be calculated from CMR and an evidence-based diagnosis of cardiovascular disease can be accomplished. A reconstruction with high spatial and temporal resolutions across the whole cardiac sequence is an indispensable prerequisite for CMR. In this context, a short scan time, ideally within a single breath-hold, is preferred to alleviate the patients' scan discomfort and prevent potential image artifacts due to patient motion. To this aim, only a limited amount of k-space (frequency domain) data can be collected for every temporal frame, violating the Nyquist–Shannon sampling theorem and resulting in aliasing artifacts in the image domain. In the past decade, Parallel Imaging~\cite{TSENSE,TGRAPPA} and Compressed Sensing~\cite{k-t-focuss,CompCSPI} were introduced in CMR, enabling shorter scan time and improved reconstruction performance. However, reconstruction performance can be further improved if adequate spatial-temporal information is shared along the cardiac cycle. This information is linked by the cardiac motion, which bridges every single frame of the whole cardiac sequence and serves as the key to successful reconstruction. A straightforward way to leverage this motion information in CMR reconstruction is to use motion-compensated MR reconstruction (MCMR)~\cite{Batchelor2005} in which the cardiac motion has to be estimated. However, precise cardiac motion estimation remains a challenging problem due to the non-rigid nature of the cardiac motion, especially in the case of accelerated imaging where motion has to be estimated from undersampled data.

 \textbf{\textit{CMR reconstruction.}} To circumvent the non-trivial tasks of cardiac motion estimation, different CMR reconstruction methods sidestep the motion estimation and aim to exploit spatio-temporal redundancies. The works of \cite{LplusS,Huang21} suggested disentangling the original reconstruction problem into a low-rank and a sparse component and these two sub-optimizations are carried out jointly. However, the preservation of dynamic information crucially depends on the optimization of the sparse component and the implementation of soft thresholding can incur information loss. Moreover, deep learning reconstructions were proposed e.g. \cite{Kuestner2020CINENet,dl-espirit} that unroll the dynamic MR optimization process with a spatio-temporal regularization. In this case, multiple unrolled gradient descent steps have to be executed, giving rise to the training difficulty of the network and processing time in both training and testing. Other methods \cite{kt_slr,Qin2019,CTFNet} utilized the $k-t$ domain to leverage the spatio-temporal redundancies to ameliorate the dynamic reconstruction or resolve the motion implicitly after the image acquisition~\cite{ghodrati2021retrospective,kustner2019retrospective}. Whereas all these methods endeavor to extract the spatio-temporal correlation implicitly, there is no guarantee that the correlation of every cardiac frame is fully exploited. On the contrary, MCMR leverages the estimated cardiac motion to explicitly share cardiac spatio-temporal information. %The motion estimation performance can be visually represented by color wheel~\cite{colorwheel} or quiver and evaluated quantitatively or qualitatively. 

 \textbf{\textit{Motion estimation in MCMR.}} A high-quality MCMR can be performed if the cardiac motion can be estimated precisely over the whole cardiac cycle. Therefore, the selection of a proper motion estimation/registration approach plays a decisive role in MCMR. Conventional registration methods based on B-spline~\cite{klein2009elastix,Modat2010} or diffusion method~\cite{vercauteren2009diffeomorphic} can be employed as motion estimators in MCMR. These methods can provide meaningful registration results but demand enormous computing time in the order of hours for a single CMR sequence. Furthermore, hyperparameter tuning for these methods~\cite{klein2009elastix,Modat2010} is also a non-trivial task, hindering their implementation in clinical practice. Lately, learning-based registration/motion estimation approaches have been introduced into medical imaging~\cite{registration_review,voxelmorph1,voxelmorph2} and embodied in the application of cardiac motion estimation~\cite{Morales2019,MPN2020,Pan2021}. These methods accelerate the registration time from hours to seconds by leveraging a trained neural network during inference and mitigating hyperparameter tuning. However, these cardiac registration methods are not designed for the MCMR context but are designed to minimize the brightness inconsistency of estimated motion-warped images and target images (motion-warping error). Yet in the context of accelerated imaging, the undersampled input images exhibit artifacts and intensity inconsistencies. The direct application of these general motion estimation/registration methods to accelerated imaging data can result in imprecise motion fields and can thus incur error propagation in MCMR. Qi \emph{et al.}~\cite{Qi2021-2} circumvented this problem by providing reference images in the training loss whilst feeding undersampled data as network inputs. Concurrently, a registration method designed for the MCMR context is proposed by Küstner \emph{et al.}~\cite{LAPNet} in which the registration is directly estimated from the k-space. All aforementioned methods conduct a pair-wise motion estimation and they have to be carried out multiple times in MCMR, in which for every single frame a registration from multiple other frames is required. To provide a more efficient and time-continuous registration, group\-wise motion estimation has been studied~\cite{Hammernik2021,Qi2022}. In group-wise registration, the spatial-temporal redundancy over multiple frames can be leveraged to facilitate the registration, especially when through-plane motion occurs in the context of 2D CMR. Furthermore, the temporal coherence over the cardiac cycle can be instilled during training by applying a temporal loss term~\cite{Hammernik2021}.

 \textbf{\textit{MCMR framework.}} After the choice of a proper motion estimation/registration method, there are multiple MCMR frameworks available to combine the motion estimator and reconstruction. The seminal work~\cite{Batchelor2005} of Batchelor \emph{et al.} pioneered the MCMR concept in which the motion information is embedded as a general matrix into the MR forward model. This work formulated the MCMR problem with two individual stages: motion estimation and reconstruction. The motion estimation in the first stage and the reconstruction in the second stage are both carried out separately, while the pre-calculated motion from the first stage is regarded as a fixed matrix in the second stage reconstruction. This two-stage approach has been widely used for respiratory motion compensated reconstruction~\cite{Prost1,Prost2,Qi2021-2,MUNOZ202210,zou2022dynamic} but rarely for cardiac CINE~\cite{Cruz2021}. Furthermore, MCMR can also be reformulated as a joint optimization problem in which an iterative optimization of image reconstruction and motion estimation are carried out alternatively. A potential synergy can be established: a more accurate motion estimation can provide a better reconstruction, and based on a less artifacts-affected image a better motion estimation can be accomplished. Odille \emph{et al.} proposed a reconstruction method using sensor-based motion estimation e.g. respiratory belt or ECG signal~\cite{Odille2008,Odille2016}. The need for external tracking hardware is relieved by adopting B-spline-based and optical flow-based motion estimation in this joint optimization context~\cite{NiftyReconJoint,FlowReconJoint}. More recently, variational methods \cite{CS_M} and dictionary learning~\cite{CS_M2} are also employed to solve this joint optimization problem for CMR reconstruction. However, all these methods demand a relatively long estimation time because of their iterative optimization nature. Therefore, deep-learning-based methods were proposed to speed up joint optimization. \cite{Qi2022,Pan2022-1} unrolled MCMR joint optimization with a group-wise motion estimation network and the mutual benefit of CMR reconstruction and motion estimation is demonstrated in their work. % \cite{Seegoolam} proposed cascaded neural networks to estimate the motion and perform reconstruction gradually with incremental usage of cardiac frames.

 \textbf{\textit{Drawbacks of decomposition.}} However, the decomposition of the MCMR into two sub-optimization problems serves as a workaround to solve MCMR has two major drawbacks: First, the solution space of the full problem is restricted by the solution of the motion-estimation problem itself whose goal is to minimize the motion-warping loss between different cardiac frames. This goal is not necessarily aligned with the final reconstruction objective due to undersampled images' artifact-degradation and intensity-inconsistency amongst cardiac frames. Second, extra efforts have to be built in to cope with motion estimation in the case of accelerated imaging with undersampled data, e.g. extra pre-processing steps with intra-bin motion correction~\cite{Prost1,Prost2}, loss function tuning~\cite{Qi2021-2} or k-space motion estimation~\cite{LAPRecon}. Although the estimation difficulty of the motion can be reduced if the alternating joint optimization is used, it requires multiple iterations of motion estimation and reconstruction to yield satisfactory reconstruction, prolonging the processing time. On the contrary, in this work we propose an MCMR framework that optimizes the complete MCMR framework together without breaking it into two sub-optimization problems.

% However, splitting the MCMR into these two separate processes imposes high demands on the motion estimator to be accurate, especially in the case of accelerated imaging with undersampled data. Some works applied extra pre-processing steps like intra-bin motion correction to mitigate the motion estimation burden~\cite{Prost1,Prost2,Qi2021-2}. Besides, predicting the motion from k-space data can also facilitate a two-stage MCMR \cite{LAPRecon}. If deep learning is adapted in MCMR, the utilization of fully-sampled images instead of undersampled images in the training loss function can also boost the motion estimation and reconstruction quality~\cite{Qi2021-2}.

Moreover, all aforementioned MCMR methods follow the suggestions of~\cite{Batchelor2005} which applied all temporal frames to reconstruct one single frame of the sequence so that all temporal redundancy can be exploited. We argue in this work that using a smaller amount of temporal frames to conduct the MCMR can achieve a better result. This setting reduces the residual motion-warping error from other temporal frames while still leveraging enough redundant information. 
%Therefore, in this work, we study the impact of a smaller number of temporal frames to reconstruct every cardiac frame and find using a smaller amount can achieve better reconstruction results than using all.

% Moreover, all aforementioned MCMR methods which follow the optimization procedure proposed by~\cite{Batchelor2005} apply all temporal frames of the sequence to reconstruct the individual frame. %However, the motion-warping error cannot be suppressed to zero even with the state-of-the-art motion estimation method. 
% We argue that using a smaller amount of temporal frames to conduct the MCMR can achieve a better result which can reduce the residual warping error from other temporal frames while still leveraging enough redundant information.

\begin{figure*}
	\centering
	\includegraphics[width=1\textwidth]{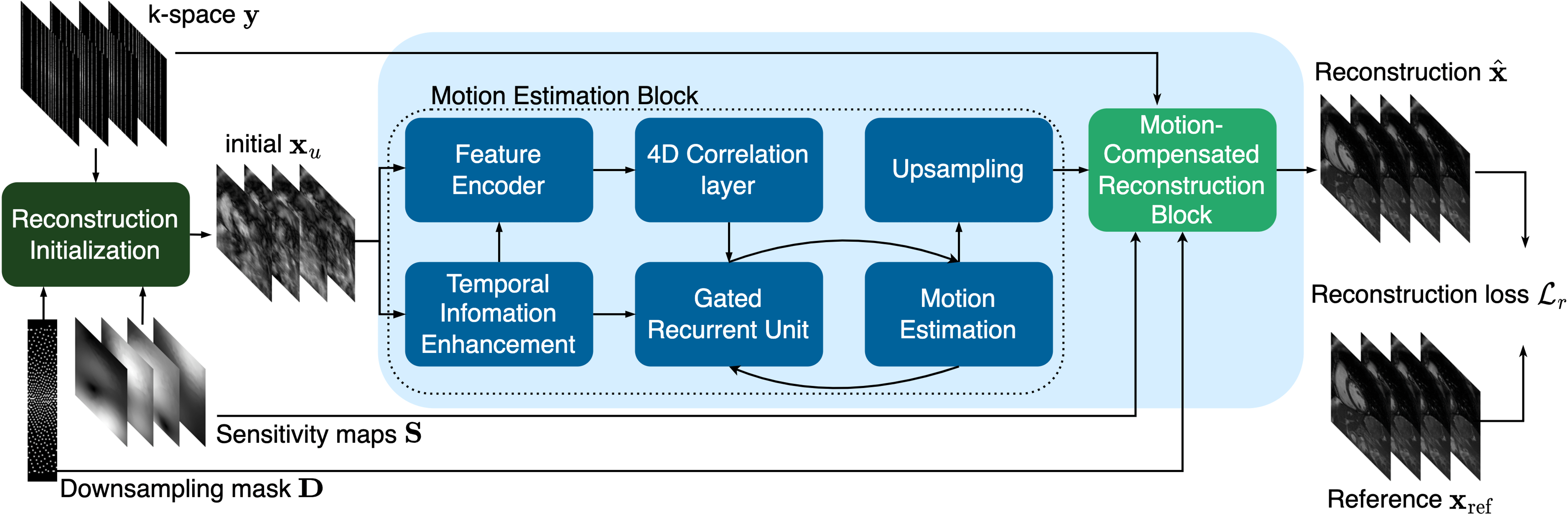}
	\caption{Architecture of the proposed method: Motion-compensated MR reconstruction (MCMR) framework with a \textit{Motion Estimation Block} (refer to ~\ref{motion}) and a complex-valued \textit{Motion-Compensated Reconstruction Block} (refer to ~\ref{recon}). The motion estimation learning process is directly driven by the final reconstruction performance. A pre-processing reconstruction is implemented (\textit{Reconstruction Initialization}, refer to ~\ref{CGSENSE}) prior to the proposed method to alleviate the reconstruction difficulty.}
	\label{fig1}
\end{figure*}

In summary, the main contributions of our work are as follows: 
\begin{enumerate}
    \item We propose a deep learning-based approach, which efficiently solves the motion-compensated reconstruction and addresses the MCMR problem as a single entity. Our framework estimates motion from the perspective of CMR reconstruction, rather than motion estimation alone. We establish an efficient mechanism in which the motion estimation process is directly driven by the final reconstruction results (refer to Fig.~\ref{fig_overview}) and without using iterative joint optimization of motion estimation and reconstruction. 
    \item We investigate the optimal number of temporal frames to use during the MCMR. We observe that using a smaller amount of frames to reconstruct the cardiac frames achieves better performance than using all frames of a sequence. We find a balance between the exploitation of sequence redundancy and the suppression of residual warping error, which can inspire all other MCMR methods. 
    \item We demonstrated the reconstruction of images from undersampling rates up to 20x with the optimization depending on \textbf{only one} final reconstruction loss term. The canonical motion-warping loss including regularization/smoothness terms that serve as an intermediate loss in MCMR is discarded in this work. Therefore, we avoid the non-trivial weighting factors tuning.
    \item We applied our method on in-house acquired CMR CINE data for the \textbf{retrospective} study and also on the public OCMR cardiac data \cite{chen2020ocmr} for the \textbf{prospective} study. We compare our method to several canonical and SOTA methods. The proposed method outperforms the baselines in both qualitative and quantitative evaluation with more superior and robust performance.
\end{enumerate}

\section{Problem Formulation}
\subsection{General MR Reconstruction}\label{CGSENSE}
Let $x^{(n)} \in \mathbb{C}^{M}$ indicate the $n$-th complex-valued temporal frame of the dynamic CINE sequence $\mathbf{x} =  [x^{(1)},\ldots,x^{(N)}]^T$ ("," indicates vector appending) stacked as a column vector and $M$ denotes the number of pixels in the 2D plane, i.e. $M=M_{X}M_{Y}$ with $X$, $Y$ the height and width of the frame and $N$ the number of temporal phases. $y^{(n)} \in \mathbb{C}^{SM}$ from $\mathbf{y} =  [y^{(1)},\ldots,y^{(N)}]^T$ is the corresponding undersampled k-space data with $S$ being the number of MR receiver coils. Regarding the CMR reconstruction task of a retrospectively gated CINE, the following inverse problem has to be solved:

\begin{align}
\label{eq1}
\min\limits_{x^{(n)}} \left\Vert \mathbf{A}^{(n)} x^{(n)}  - y^{(n)} \right\Vert^2_2, \ n=1,\dots, N. 
%$+ \lambda\mathcal{R}(x^{(t)}). 
\end{align}
$\mathbf{A}^{(n)}$ represents the MR forward multi-coil encoding operator with $\mathbf{A}^{(n)}=\mathbf{D}^{(n)}\mathbf{F}\mathbf{S}$, in which $\mathbf{S} \in \mathbb{C}^{SM \times M}$ denotes the coil sensitivity maps, $\mathbf{F} \in \mathbb{C}^{SM \times SM}$ is the forward Fourier encoding matrix, $\mathbf{D}^{(n)} \in \mathbb{R}^{SM \times SM}$ is the undersampling mask diagonal matrix. The value on the diagonal is 1 if the corresponding k-space data is sampled, and 0 if unsampled. %and the subscript $g$ denotes "general". $\mathcal{R}$ presents the regularization term upon $\mathbf{x}$ and $\lambda \in \mathbb{R}$ denotes the weighting factor to balance these two terms. 
Eq. \eqref{eq1} can be solved by using general conjugate-gradient SENSE (CG-SENSE)~\cite{Pruessmann2001} reconstruction which is performed $N$ times to reconstruct these $N$ cardiac frames. However, this general MR reconstruction method optimizes every cardiac frame $x^{(n)}$ separately regardless of the adequate temporal information across the cardiac sequence. Therefore, its reconstruction performance is limited with respect to the undersampling ratio. In this work, we use this general CG-SENSE as an initialization step (\textit{Reconstruction Initialization} in Fig.~\ref{fig1}) to facilitate the following MCMR task. It is to note that we do not carry a complete CG-SENSE here but an optimization limited to 10 iterations. The optimal reconstruction during the initialization step is attained after approximately 300 iterations, resulting in a PSNR enhancement of 0.5dB in the final performance. However, this comes at the expense of a significantly extended reconstruction duration. By terminating the optimization at the 10th iteration, we achieve a balance between efficiency and speed in the initialization stage.

\subsection{Motion-compensated MR reconstruction with a varying number of input neighboring frames}
As mentioned above, leveraging the temporal information in the cardiac sequence can facilitate the CMR reconstruction. The spatial-temporal redundant information is bridged by the cardiac motion. Following the work of Batchelor \emph{et al.}~\cite{Batchelor2005}, motion is embedded into the MR forward model and information from other temporal frames can be leveraged as complements:

% block matrix formulation of MCMR
\begin{align}
\label{eq2}
\min\limits_{x^{(n)}} \left\Vert \mathbf{A}^{(K)} \mathbf{U}^{(n \rightarrow K)}x^{(n)}  - \mathbf{y}^{(K)} \right\Vert^2_2,\ n=1,\dots, N 
\end{align}
where $K=2k+1$ denotes the neighboring $\pm k$ frames of the frame $n$. The k-spaces $\mathbf{y}^{(K)} = [y^{(n-k)},\dots, y^{(n)},\dots,y^{(n+k)}]^T \in \mathbb{C}^{SMK}$ are used as complementary neighboring data to reconstruct the frame $x^{(n)}$. We assume periodicity in the cardiac cycle, i.e. the previous frame to $x^{(0)}$ is regarded as $x^{(N)}$. $\mathbf{U}^{(n\rightarrow K)} \in \mathbb{R}^{MK \times M}$ denotes the cardiac motion matrix and warps $x^{(n)}$ to the $K$ cardiac frames. By means of $\mathbf{U}^{(n \rightarrow K)}$, the redundancy and correlation of the neighboring cardiac frames of $x^{(n)}$ are instilled for the $n$-th frame reconstruction. It should be noted that our MCMR framework differs from the original MCMR framework~\cite{Batchelor2005} which applied all temporal frames $K=N$ to conduct the reconstruction, while in our case we choose $K<N$ as detailed in Section~\ref{ablation}. Analogously to $\mathbf{A}^{(n)}$, $\mathbf{A}^{(K)}=\mathbf{D}^{(K)}\mathbf{F}\mathbf{S} \in \mathbb{C}^{SMK \times MK}$ denotes the CMR forward model for these $K$ frames.

\section{Method}
In this work, we propose a deep-learning-based framework to reconstruct the dynamic CINE images. This framework consists of two parts: a \textit{Motion Estimation Block} which tries to estimate cardiac motion and a \textit{Motion-Compensated Reconstruction Block} which is purposed for carrying out the motion-compensated reconstruction, depicted in Fig.~\ref{fig1}. In contrast to all the previously proposed MCMR works, our framework can be trained end-to-end, which regards the motion estimation and reconstruction processes as a single entity instead of splitting them into two sub-tasks. Furthermore, unrolling the iterative procedure of motion estimation and reconstruction prolongs the processing time and renders itself inefficient. In this work we aim to estimate precise motion directly from the undersampled data by using one-shot prediction with a motion estimator $\mathcal{G}$ and then solve the inverse problem with an $\ell_2$ regularizer using the initial sequence $\mathbf{x}_u$ provided by \textit{Reconstruction Initialization} block, read as:
% i.e. Eq. 4 is only performed once. Therefore, Eq.~\eqref{eq4} and Eq. 4b are recast as:

\begin{subequations}
\begin{align}
\label{eq6}
\hat{\mathbf{U}}&=\mathcal{G}\left(\mathbf{x}_u\right) \\
\hat{\mathbf{x}} &=\arg \min _{\mathbf{x}}\left\|\mathbf{A} \hat{\mathbf{U}} \mathbf{x}-\mathbf{y}^{(NK)}\right\|_2^2+\lambda\left\|\mathbf{x}-\mathbf{x}_u\right\|_2^2, \label{eq6b}
\end{align}
\end{subequations}

where $\lambda$ presents the weighting factor of the $\ell_2$ term, which is commonly used in MR reconstruction~\cite{sandino2020compressed}. The regularizer here is a variant version of $\ell_2$ regularization, which can improve the conditioning of the problem and adjust the temporal resolution over the cardiac cycle. \textbf{However, we find that in our work, this regularization term is not obliged to apply. We can still achieve satisfactory performance without using it. Therefore, we can discard this term and set $\lambda$ to 0}. More details can be found in \ref{abs} and in our ablation study \ref{ablation}. $\hat{\mathbf{U}} \in \mathbb{R}^{MNK \times MN}$ denotes the estimated cardiac motion which is used inside the MCMR, $\mathbf{A}=\mathbf{D}^{(NK)}\mathbf{F}\mathbf{S} \in \mathbb{C}^{SMNK \times MNK}$ indicates the forward model for these $NK$ frames, $\hat{\mathbf{x}}$ denotes the final reconstructed image for all cardiac frames. $\mathbf{y}^{(NK)} = [y^{(1-k)},\dots,y^{(1+k)},\dots\dots,y^{(N-k)},\dots,y^{(N+k)}]^T$ extends from $\mathbf{y}^{(K)}$ presenting the adopted complementary neighboring frames to reconstruct every $x^{(n)}$ of the sequence $\mathbf{x}$.

\subsection{Motion Estimation Block}\label{motion}
We utilize a learning-based motion estimation network $\mathcal{G}$ with trainable parameters $\theta$ to predict the non-rigid cardiac motion. The backbone of GRAFT~\cite{Hammernik2021} is applied to model $\mathcal{G}_{\theta}$. The choice of using GRAFT rather than other registration methods is anchored in its demonstrated registration accuracy, speed and efficiency, as reported in our previous work~\cite{Pan2022-1}. GRAFT is a group-wise motion estimation network that takes the undersampled cardiac sequence $\mathbf{x}_u$ as input and predicts the motion between the frames. Its inherent \textit{Temporal Information Enhancement} Block consists of convolutional layers that take the target frame along with its one previous and subsequent cardiac frame as input and extract the spatial-temporal information from them. By means of that, the problem of through-plane motion and occlusion can be alleviated. Afterward, a \textit{Feature Encoder} is incorporated which processes the embedding from \textit{Temporal Information Enhancement Block} and extracts the meaningful features for the motion estimation from the image sequence. Subsequently, a \textit{4D-Correlation layer} is performed to compute the correlation of the 2D spatial planes. A global correlation search between the feature embedding of the two cardiac frames is conducted within this layer, which attempts to capture not only small but also large deformation. A \textit{Gated Recurrent Unit (GRU)} is employed afterward to conduct an iterative motion estimation. This process can be regarded as an optimization procedure in which the estimation is refined and the residual error is removed over the course of the iterations gradually. Finally, the motion is upsampled $4\times$ to the original image size. Importantly, we do not use linear interpolation but implement a learnable convolutional upsampler to upscale the motion more precisely as suggested by~\cite{teed2020raft}. This process is carried out $K$ times and a motion field $\hat{\mathbf{U}}$ mapping from dimension $MN$ to $MNK$ is produced by GRAFT at this end. 

Usually, a warping similarity measurement $\mathcal{L}_w$ is utilized to drive the learning of the motion estimation network: $\mathcal{L}_w(\mathbf{x}^{(K)}, \hat{\mathbf{U}}^{(n\rightarrow K)}x^{(n)})$ with $\mathbf{x}^{(K)} = [x^{(n-k)},\dots,x^{(n)},\dots,x^{(n+k)}]^T$ the target frames from $\mathbf{x}$ and $\hat{\mathbf{U}}^{(n\rightarrow K)}x^{(n)}$ its corresponding warped estimation. However, $\mathcal{L}_w$ is just an intermediate motion-warping loss function in the context of MCMR. As mentioned in Section~\ref{sec:introduction}, the loss' effectiveness is undermined by the increase of the undersampling rate (more aliasing and severe intensity inconsistency) whose goal diverges from the goal of improving the final reconstruction quality. Furthermore, the utilization of $\mathcal{L}_w$ after the \textit{Motion Estimation Block} breaks the original MCMR optimization into two sub-tasks, introducing the drawbacks as mentioned in Section~\ref{sec:introduction}. In this work, we do not calculate $\mathcal{L}_w$ at this intermediate position but forward the output motion $\hat{\mathbf{U}}$ of $\mathcal{G}$ to the subsequent \textit{Motion-Compensated Reconstruction Block}. Since no network loss function is applied yet, the motion prediction $\hat{\mathbf{U}}$ with learnable parameters $\theta$ are still pending and the complete forward chain of the applied deep learning model is to be established by the subsequent \textit{Motion-Compensated Reconstruction Block}.

\subsection{Motion-Compensated Reconstruction Block}\label{recon}
The \textit{Motion-Compensated Reconstruction Block} is a complex-valued operator that executes the CINE reconstruction. It endeavors to solve Eq. \eqref{eq6b} by finding the stationary point utilizing the normal function:
\begin{align}
\label{eq9}
\underbrace{\left(\hat{\mathbf{U}}^H\mathbf{A}^H\mathbf{A}\hat{\mathbf{U}}+\lambda \mathbf{I}\right)}_{\mathcal{V}}\hat{\mathbf{x}}=\underbrace{\left(\hat{\mathbf{U}}^H\mathbf{A}^H\mathbf{y}^{(NK)}+\lambda\mathbf{x}_u\right)}_{\mathbf{b}}.
\end{align}
The inverse of matrix $\mathcal{V}$ is computationally prohibitive to calculate. Inspired by MoDL~\cite{MoDL}, Conjugate Gradient (CG)~\cite{CG} is adopted and wrapped in this work within this \textit{Motion-Compensated Reconstruction block}. CG solves Eq.~\ref{eq9} in an iterative manner until the process converges. We fix the number of iterations as $I$ and present this CG-algorithm-based block as a mapping function $\mathcal{F}$. It takes variable $\hat{\mathbf{U}}$ and the optional hyperparameter $\lambda$ as inputs and yields reconstruction $\hat{\mathbf{x}}_I$ as output, which reads as: 
\begin{align}
\label{eq10}
\hat{\mathbf{x}}_I = \mathcal{F}(\hat{\mathbf{U}}, \lambda).
\end{align}
\textbf{In contrast to conventional deep learning-based MCMR works there are two major differences of the proposed method we want to emphasize here.} First, it is important to note that $\hat{\mathbf{U}}$ here is not a static fixed matrix but still a pending variable from $\mathcal{G}_{\theta}(\mathbf{x}_u)$ without gradient-stop. Its gradient and trainable parameters $\theta$ still wait for updates through back-propagation on a higher-level loss function for network training. Second, the \textit{motion-Compensated Reconstruction block} i.e. mapping function $\mathcal{F}$ itself does not have any trainable parameters but serves as the forward pass for $\hat{\mathbf{U}}$ to reach the final loss function. To express Eq. \eqref{eq10} more clearly, we reformulate it as $\hat{\mathbf{x}}_I = \mathcal{F}(\mathcal{G}_{\theta}(\mathbf{x}_u), \lambda).$

Finally, we define our loss function $\mathcal{L}_r$ as the mean squared error between the reconstruction estimation $\hat{\mathbf{x}}_I$ and the reference reconstruction target $\mathbf{x}_\text{ref}$. Thus, the final learning-based optimization function can be represented as:

\begin{align}
\label{eq12}
\mathcal{L}_r = \left\Vert \mathcal{F}\left(\mathcal{G}_{\theta}(\mathbf{x}_u), \lambda\right) - \mathbf{x}_\text{ref}\right\Vert_2^2.
\end{align}

Now, the complete deep-learning forward chain is established and $\theta$ can be updated by gradient back-propagation. An end-to-end MCMR framework is cast without employing any intermediate motion-warping loss. In this respect, the motion estimation process is directly guided and driven by feedback from the final reconstruction performance but not by the motion estimation/registration. The goal of motion compensation is now aligned with the final reconstruction goal.
%The spatio-temporal sharing of samples in the final motion-compensated images is now aligned. 

\section{Experiments}
\subsection{Dataset}\label{sub_A}
Two datasets are applied in this work: one is in-house acquired cardiac CINE data for retrospective reconstruction study and the other is OCMR dataset~\cite{chen2020ocmr} for the prospective study. The experiments are mainly performed in the retrospective study to train and investigate the performance and the effectiveness of the proposed method with the aid of the groundtruth images. The prospective study is carried out in inference without any fine-tuning to testify the robustness and generalizability of the method.

\paragraph{\textbf{In-house acquired CINE data}} 43 subjects (27 patients and 16 healthy volunteers) were scanned with a 2D cardiac CINE sequence. The data is acquired on a 1.5T MRI scanner (Magnetom Aera, Siemens Healthineers) with an acquisition sequence of 2D balanced steady-state free precession (bSSFP) equipped with a multi-channel body and spine coil.  Depending on the field of view placement and size, the coil channels are automatically selected, resulting in 30, 34 or 38 MR receiver coil channels. A $2\times$ GRAPPA acceleration generated the CINE data with an in-plane resolution of $1.9\times1.9 \text{mm}^2$, a slice thickness of 8mm, echo time (TE) of 1.06ms, and repetition time (TR) of 2.12ms. Retrospective gating is used to bin the data into 25 cardiac phases with a temporal resolution of 40ms. Matrix size varies from the smallest size 176 (frequency-encoding) $\times$ 132 (phase-encoding) to the largest size 192 $\times$ 192. An amount of 10 to 15 short-axis slices for each subject (stacked along the long-axis) was acquired from base to apex under multiple breath-holds (2 slices per breath-hold). Slices without clear cardiac anatomy were discarded, resulting in a total of 366 cardiac motion-resolved image sequences. Retrospective undersampling is performed by Cartesian VISTA ~\cite{Ahmad2015} sampling with varying acceleration factors for both training and inference.

\paragraph{\textbf{OCMR CINE data}} We utilized the OCMR CINE data acquired from a 1.5T Siemens Magnetom Avanto scanner, which is prospectively undersampled from 8 healthy volunteers using a short-axis plane. The data is collected with 18 receiver coils using a bSSFP sequence. The acquisition is conducted in the real-time mode under free-breathing conditions with VISTA sampling mask and the acceleration rate of $R=9$. No respiratory motion correction was applied assuming shallow breathing and given that it is a single-slice “real-time”-like acquisition with 4 heartbeats as reported in their previous work~\cite{chen2019sparsity}. Their in-plane resolution varies from $2.0\times2.0 \text{mm}^2$ (smallest) to $2.3\times2.3 \text{mm}^2$ (largest) with a slice thickness of 8mm, while the matrix size varies from the smallest size 160 (frequency-encoding) $\times$ 120 (phase-encoding) to the largest size 192 $\times$ 140. The temporal resolution varies from 37ms to 41ms with 64 frames. Slices without clear cardiac anatomy were discarded, resulting in 15 CINE sequences. These sequences are used as qualitative evaluation in inference after training on the in-house acquired cardiac CINE data. The ESPIRiT algorithm~\cite{uecker2014espirit} is employed to estimate the coil sensitivity maps. More details of the applied data and its acquisition can be found in~\cite{chen2020ocmr,chen2019sparsity}.

\subsection{Implementation Details}
The proposed framework was implemented in PyTorch (v1.9.0) and trained on an NVIDIA A40 GPU. The AdamW~\cite{adamw} optimizer combined with a one-cycle learning rate scheduler (max. learning rate 0.0001) was used to optimize Eq. \eqref{eq12}. The network parameters for the \textit{Motion Estimation Block} follows~\cite{Hammernik2021}. The hyperparameter $I$ is set to $10$ for training. It can be adapted flexibly in inference and runs until the saturation of the data consistency cost. However, every iteration in CG-SENSE costs around 0.16s on our hardware and we found $I=10$ as the "sweet spot" regarding execution time and accuracy. Based on our observations, the performance lifting after the 10th iteration is limited. Therefore, to keep our method fast in inference, we set $I$ to 10 in inference for the following experiments. Regarding network training, we adopt either a fixed undersampling rate for training or a mixed training procedure with $R=8, R=12, R=16$ and $R=20$ undersampled data with a random selection with the same probability (dubbed as mixed $R$ training). During inference, we can test our approach on an arbitrary undersampling rate for the retrospective study. The undersampled raw k-space data is first reconstructed by the \textit{Reconstruction Initialization} block and then fed to the proposed network. 

\subsection{Ablation study}\label{abs}
% relocated
\paragraph{\textbf{Amount of neighboring frames used in MCMR}} We first investigate the impact of using different amounts of neighboring frames $K$ for the dynamic CINE reconstruction during training and test. In the ideal case, the motion across the whole cardiac cycle can be estimated precisely, therefore all $N$ temporal frames should be used to exploit temporal redundancy. However, non-rigid contraction and expansion of the heart are challenging to estimate and given the 2D acquisition nature through-plane motion and occlusion (especially towards basal slices) can occur. Thus, the residual frame-to-frame warping error cannot be suppressed completely to zero even with SOTA motion estimators. If more neighboring frames are considered, a larger accumulated residual motion error and an averaging effect of the cardiac cycle can occur. We therefore investigate the optimal number of neighboring frames to use for the CINE reconstruction. We run experiments using neighbouring $k=\pm 1$ ($K=3$), $\pm 2$, $\pm 4$, $\pm 6$, $\pm 8$, $\pm 12$ ($K=25$) frames with mixed $R$ training and test on different acceleration rates. 

\paragraph{\textbf{Sensitivity of $\lambda$ and its interplay with $K$}}
The presented regularization term in Eq.~\ref{eq6} has two purposes. First, it can improve the conditioning of the problem by adding the positive elements to the diagonal (as shown on the left-hand side of Eq.~\ref{eq9}), decreasing the matrix condition number and improving stability. Second, it can adjust the temporal resolution. In the case of underestimated (or not captured) large or through-plane motion underestimated, the target frame is averaged with the other temporal frames. By introducing $\mathbf{x}_u$ into the regularizer, it can encourage the output to be close to $\mathbf{x}_u$ (in which every temporal frame differs from each other) and encourage temporal diversity. However, it also raises a trade-off because the final solution would also be biased by $\lambda\mathbf{x}_u$ (the right-hand side of Eq.~\ref{eq9}). We carry out experiments to investigate the sensitivity of this regularizer and its interplay with $K$. The right choice of $K$ can also suppress the estimation error and minimize the averaging effect of the cardiac cycle to ensure the temporal resolution, rendering the regularization term non-essential.

\paragraph{\textbf{Training strategy and loss functions}} We further investigate the benefits of using Eq.~\eqref{eq12} as the loss function in comparison to the widely used motion-warping loss $\mathcal{L}_w$ (refer to~\ref{motion}) which breaks the MCMR into two sub-tasks for the MCMR reconstruction. In addition, we employ the motion estimated from fully-sampled images (assumed to be the most precise one) as our reference motion and use it on the undersampled CMR for MCMR recovery. We use this setting to indicate the potential performance drop-off and how close our approach is to this "ideal" reference setting. To this respect, we conduct a set of experiments in which 4 trainings are carried out: 
\begin{enumerate}
    \item Training/test only uses $\mathcal{L}_r$ (Eq.~\eqref{eq12}) at acceleration rates $R=4,8,12$.
    \item Training/test only uses $\mathcal{L}_w$ at $R=4,8,12$. While the inputs of the framework are the undersampled images, in the training loss $\mathcal{L}_w$ we use reference images instead of the undersampled images as suggested by~\cite{Qi2021-2,Pan2022-1} to mitigate being affected by aliasing artifacts.
    \item We use a combined loss function $\mathcal{L} = \alpha \mathcal{L}_r + \beta\mathcal{L}_w$ (only for this ablation study). We conduct three training/test with $\alpha=10, \beta=1$ at $R=4,8,12$.
    \item "Ideal" setting using reference motion: we use $\mathcal{L}_w$ to train our motion estimator in fully-sampled ($R=1$) CMR, and then use these precise motions on undersampled CMR for MCMR recovery in inference at $R=4,8,12$.
\end{enumerate}

\subsection{Baseline comparisons}\label{sub_E}
We compare our method with six baseline methods. Two SOTA MCMR methods are considered in which the cardiac motion is estimated explicitly prior to the reconstruction. One is GRAFT-Recon~\cite{Hammernik2021}, which applies GRAFT to predict the cardiac motion by using $\mathcal{L}_w$ loss and then conducts the follow-up reconstruction task separately. The second one is Unrolled-MCMR~\cite{Pan2022-1}, which performs an iterative unrolled joint optimization of cardiac motion estimation and reconstruction but its motion is also calculated from $\mathcal{L}_w$. Moreover, sparse MRI based on compressed sensing for parallel imaging with $\ell_1$ wavelet regularization~\cite{lustig2007sparse} (abbreviated as PICS $\ell_1$ wavelet), L+S~\cite{LplusS}, 3D MoDL extended from~\cite{MoDL} and CTF-Net~\cite{CTFNet} are adopted as non-MCMR reconstruction methods for comparison. PICS $\ell_1$ wavelet reconstructs the images with $\ell_{1}$ regularization in the wavelet domain, L+S solves the problem by leveraging the decomposed low-rank and sparse matrix, 3D MoDL uses an unrolled scheme with a 3D ($x-y-t$ plane) dealiasing network and a data-consistency term, while CTF-Net tackles the problem by exploiting the $k-t$ domain redundancy using recurrent networks.

\subsection{Evaluation}\color{black}
We apply Structural Similarity Index (SSIM)~\cite{Wang2004} and Peak Signal-to-Noise Ratio (PSNR) to evaluate the reconstruction performance quantitatively. Besides these two metrics, we also employ Learned Perceptual Image Patch Similarity (LPIPS)~\cite{LPIPS} which has been verified to be closer to human perception. Furthermore, we use a cardiac segmentation network~\cite{ChenCardiacReview} to obtain a bounding box around the heart to focus evaluation on the cardiac anatomy. An offset value of $10$ pixels is set to extend the bounding box region. All metrics (SSIM, PSNR and LPIPS) are evaluated within this heart region. Moreover, we perform a targeted quantitative evaluation only on end-systolic (ES) and end-diastolic (ED) frames using PSNR, which is crucial for extracting clinical indicators like ejection fraction. The quantitative evaluation cannot be performed on the prospectively undersampled data due to the lack of fully-sampled groundtruth images. Further, we also carry out the qualitative evaluation by visualizing the reconstructed images in both $x-y$ and $y-t$ plane. Regarding the MCMR methods, its predicted cardiac motion is also demonstrated. We visualize the error maps between the reconstructed images and the reference images in the retrospective undersampling as well.

% It also adopts Eq \ref{8} as the guided loss function but it is to be noted that

\section{Results}
\subsection{Ablation Study}\label{ablation}
\begin{figure}[t!]
	\centering
	\centerline{\includegraphics[width=1\linewidth]{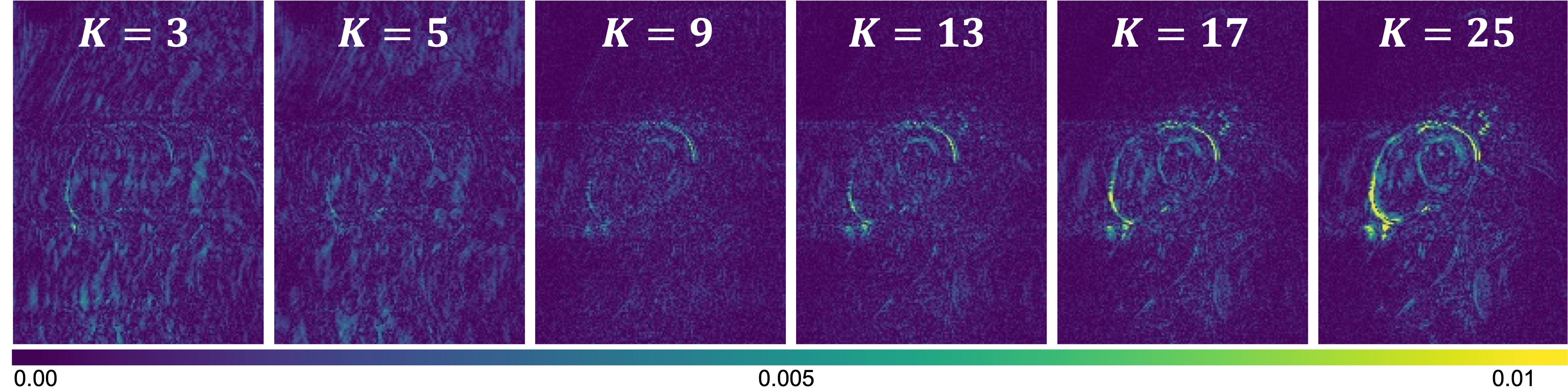}}
	\caption{Reconstruction error maps between reconstructed and reference image for using different neighboring frames amount $K$ on a test sample with acceleration rate $R=20$.}
	\label{fig2.1}
  % \vspace{-0.5cm}
\end{figure}

The ablation study is investigated in the in-house acquired CINE data.
% relocated
\paragraph{\textbf{Amount of neighboring frames used in MCMR}} We foremost delve into the study to find out the optimal number of neighboring frames for MCMR. The quantitative evaluation for the optimal amount is shown in Table \ref{t2}. A qualitative analysis of the reconstruction error for the $R=20$ case is illustrated in Fig.~\ref{fig2.1}. It can be seen that adopting fewer neighboring frames in low acceleration rates is preferable. Yet with the increase of acceleration rate for a small number of neighboring frames, no sufficient temporal redundancy is captured, resulting in increased reconstruction errors in the image background as shown in the $K=3(\pm1)$ and $5(\pm2)$ cases. However, if a high number of neighboring frames is chosen, a performance drop is observed because the advantage of using more redundant information from other frames is overcome by the suboptimal large and through-plane motion errors. In this case, motion-warping errors accumulate around the heart region while the background error is suppressed (refer to the $K=25$ case). We found that choosing $K=9$ is an optimal trade-off across all acceleration rates as indicated in Table~\ref{t2}. Therefore, we applied this value for the reconstruction in all experiments.

\begin{table}
\centering
\caption{The reconstruction performance of using different neighboring frames amount $K$. PSNR is employed here as the evaluation metric and calculated on the whole range of images. The top two results are marked in bold.}
\label{t2}
\begin{tabular}{ccccc}
\toprule
K & $R=8$ & $R=12$ & $R=16$ & $R=20$  \\
\midrule
% $3$ & 40.79 $\pm$ 2.25&37.31 $\pm$ 2.09&33.14 $\pm$ 2.44&29.52 $\pm$ 2.69\\

% $5$&\textbf{42.39 $\pm$ 2.82}& \textbf{39.73 $\pm$ 2.32} &36.73 $\pm$ 2.03&33.64 $\pm$ 2.97\\

% $9$&  \textbf{41.27 $\pm$ 3.74}& \textbf{39.85 $\pm$ 3.03}& \textbf{38.26 $\pm$ 2.57}& \textbf{36.58 $\pm$ 2.76}\\

% $13$&40.20 $\pm$ 3.72&39.16 $\pm$ 3.38& \textbf{37.92 $\pm$ 3.02} & \textbf{36.95 $\pm$ 2.83}\\

% $17$&39.31 $\pm$ 3.31&38.46 $\pm$ 3.21&37.33 $\pm$ 3.01&36.61 $\pm$ 2.99\\

% $25$&37.99 $\pm$ 2.85&37.26 $\pm$ 2.77&36.21 $\pm$ 2.69&35.77 $\pm$ 2.90\\

$3$ & 40.85 $\pm$ 2.26&37.58 $\pm$ 2.12&33.66 $\pm$ 2.42&30.08 $\pm$ 2.77\\

$5$&\textbf{42.35 $\pm$ 2.85}& \textbf{39.82 $\pm$ 2.37} &37.01 $\pm$ 2.08&34.05 $\pm$ 2.88\\

$9$&  \textbf{41.24 $\pm$ 3.71}& \textbf{39.85 $\pm$ 3.08}& \textbf{38.35 $\pm$ 2.65}& \textbf{36.71 $\pm$ 2.75}\\

$13$&40.17 $\pm$ 3.64&39.18 $\pm$ 3.38& \textbf{37.96 $\pm$ 3.05} & \textbf{37.00 $\pm$ 2.86}\\

$17$&39.30 $\pm$ 3.28&38.50 $\pm$ 3.20&37.38 $\pm$ 3.01&36.66 $\pm$ 2.98\\

$25$&38.00 $\pm$ 2.85&37.33 $\pm$ 2.78&36.31 $\pm$ 2.67&35.84 $\pm$ 2.87\\

\bottomrule
\end{tabular}
\end{table}

\begin{figure}[t!]
	\centering
	\centerline{\includegraphics[width=1\linewidth]{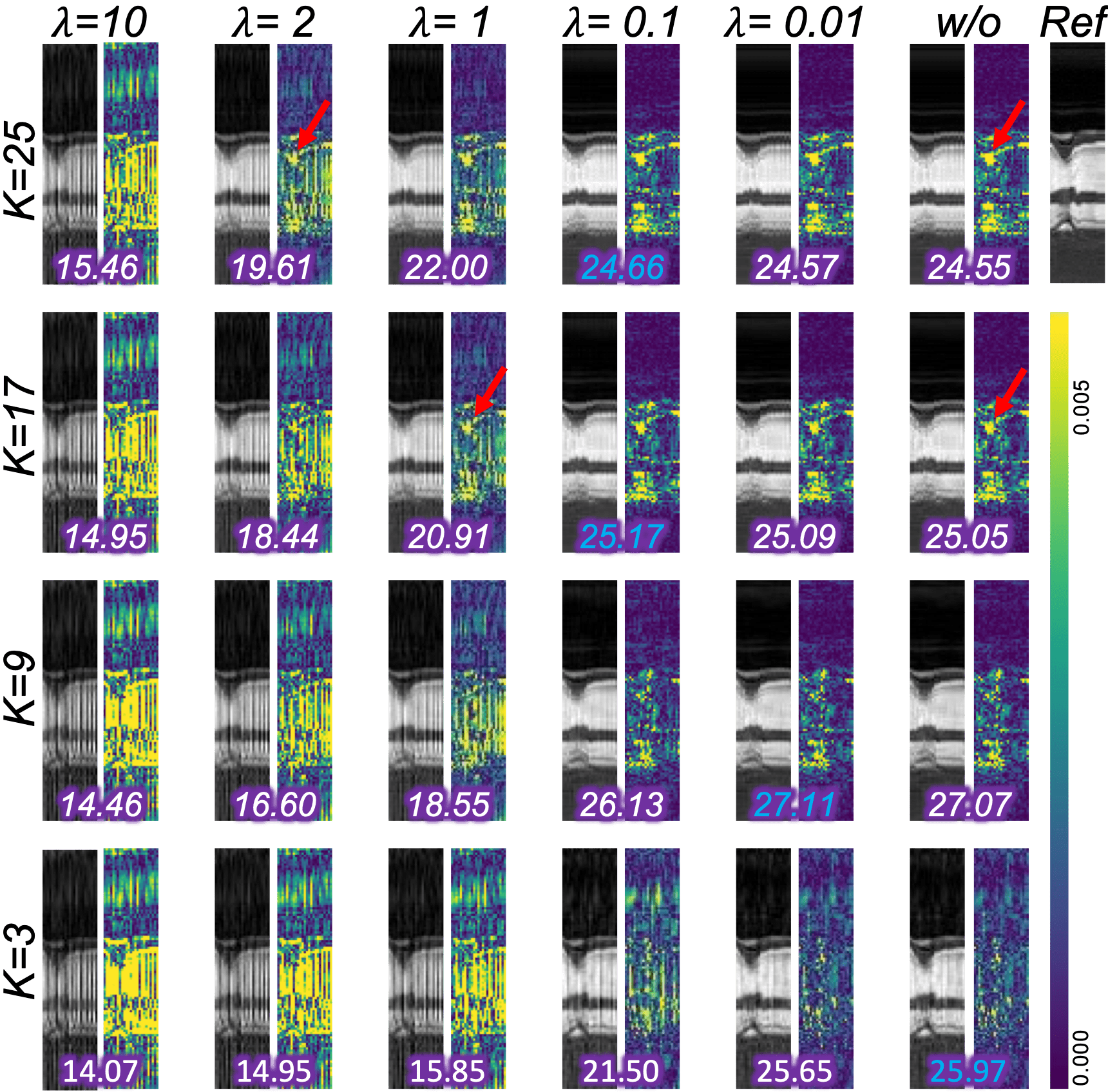}}
	\caption{The $y-t$ plane of a sample's reconstruction at $R=16$ in terms of $\lambda$ and the number of neighboring frames $K$. The corresponding averaged PSNR of the $y-t$ plane of all test samples is shown at the bottom. The best score is marked with blue. The red arrow points to larger residual errors.}
	\label{fig_lambda_k}
  % \vspace{-0.5cm}
\end{figure}

\paragraph{\textbf{Sensitivity of $\lambda$ and its interplay with $K$}}
We further analyze the effect of the optional regularization term in terms of the weighting factor $\lambda$ and the number of neighboring frames $K$. The reconstructed $y-t$ planes of one subject and the averaged PSNR of all subjects are shown in Fig.~\ref{fig_lambda_k}. The effect of the regularizer starts to kick in when $\lambda \geq 0.01$.  
The regularization term helps more if $K$ is larger. Choosing a proper $\lambda$ can help the $K=17/25$ cases elevate more than 0.1 dB in PSNR. In these cases, more frames are involved in the reconstruction and the temporal resolution is more susceptible to the underestimated motion. This effect can be visually observed if $\lambda$ increases further while the underestimated motion prediction error is compensated by the help of the regularization term with enforcement of temporal resolution (marked with red arrows in Fig.~\ref{fig_lambda_k}). However, the image quality also decreases because of the trade-off problem mentioned in~\ref{abs}. In contrast, the regularization term has limited contribution in the $K=9$ case and cannot contribute at all if $K=3$. In these cases, the temporal resolution has already been ensured by using only a few neighboring frames for reconstruction. Without this regularization, we can still achieve on-par results. Therefore in this work, we discard this regularization term and avoid the non-trivial weighting factor tuning task.

\paragraph{\textbf{Training strategy and loss functions}} We then investigate the study to find out the best training setting for MCMR and our proposed approach. The averaged PSNR of all test samples under different training strategies is shown in Table~\ref{t1}. It can be seen that the best performance of MCMR (reference) can be achieved if the motion is precisely predicted from fully-sampled ($R=1$) CMR. It is encouraged to use the reconstruction-driven loss $\mathcal{L}_r$ instead of $\mathcal{L}_w$ when images are undersampled with the occurrence of artifacts and blurring. Using $\mathcal{L}_r$ delivers superior results in all undersampled cases ($R=4,8,12$) close to the "ideal" setting. On the other hand, using conventional motion-warping loss on undersampled CINE images worsens the reconstruction. The training fails (does not converge) if only $\mathcal{L}_w$ is used since the undersampled $R=12$ images differ from the reference images severely and the motion network cannot find a correlation between them. 

\begin{figure}[t!]
	\centering
	\centerline{\includegraphics[width=1\linewidth]{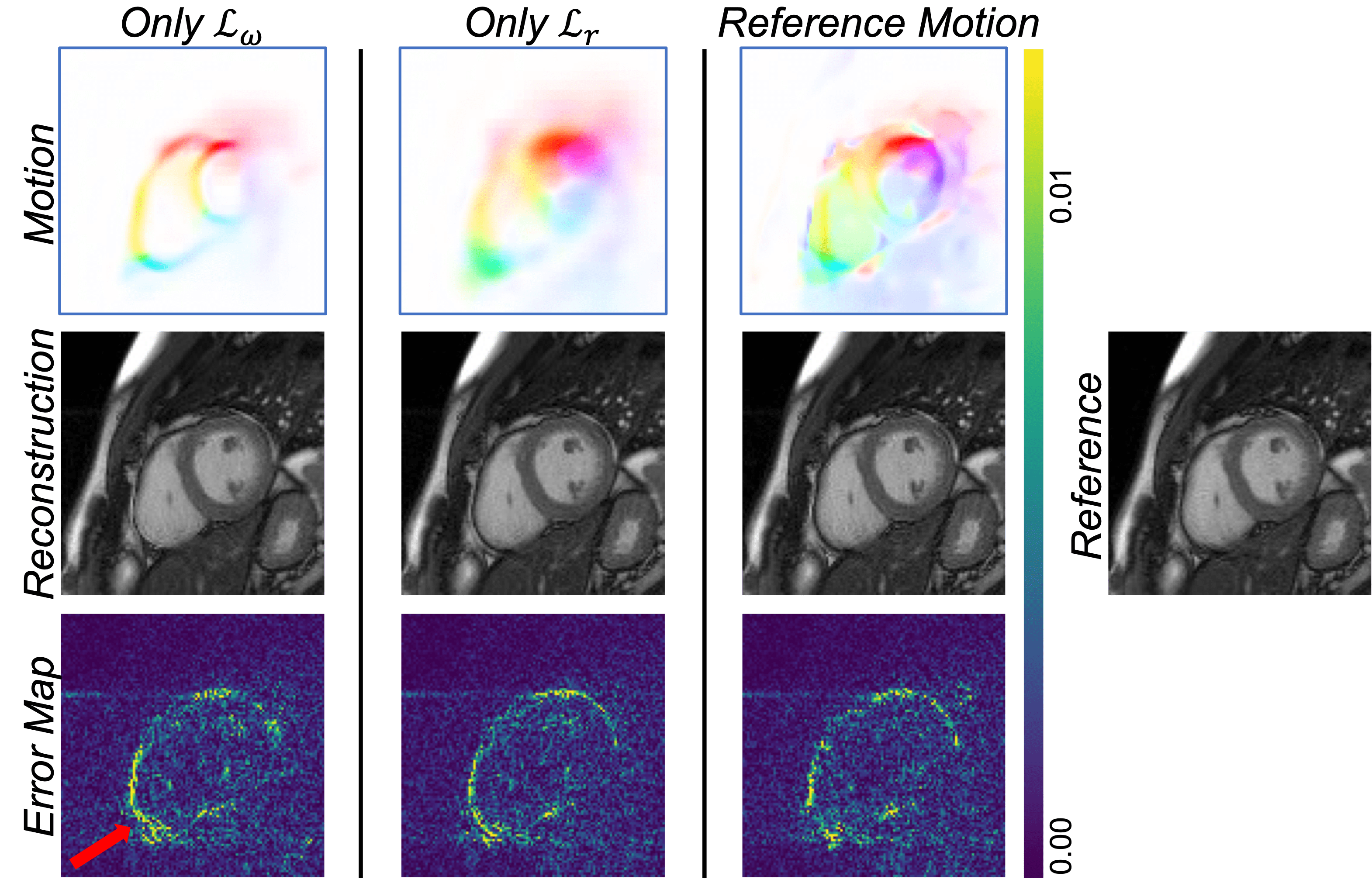}}
	\caption{The reconstruction results at acceleration rate $R=8$ with motion estimated using $\mathcal{L}_w$, $\mathcal{L}_r$ (proposed) and motion estimated from fully-sampled ($R=1$) CMR images (reference motion). The color-wheel-encoded~\cite{colorwheel} motion field, reconstructed images and the corresponding error maps are shown. The red arrow points to larger residual errors.}
	\label{figFullyUS}
  % \vspace{-0.5cm}
\end{figure}

The same conclusion of using the proposed $\mathcal{L}_r$ loss achieving superior results than using $\mathcal{L}_w$, can also be drawn from the qualitative study as depicted in Fig.~\ref{figFullyUS}. Applying the proposed $\mathcal{L}_r$ loss accomplishes a more dense and smooth motion estimation close to the reference motion. On the contrary, using the canonical $\mathcal{L}_w$ loss results in an underestimated and sparse motion field and therefore inferior reconstruction compared to the proposed method and the "ideal" setting using the reference motion.

\begin{table}
\centering
\caption{The impact of using the motion-warping loss $\mathcal{L}_w$ and the proposed end-to-end reconstruction-driven training loss $\mathcal{L}_r$ with training/test on $R=4,8,12$. Further, motion estimated using $\mathcal{L}_w$ from fully-sampled $R=1$ data and test on $R=4,8,12$ for MCMR is shown as reference. PSNR is employed here as the evaluation metric. The failed training is marked with 'N.A.'. The top two results are marked in bold.}
\label{t1}
\begin{tabular}{lcccc}
\toprule
  & only $\mathcal{L}_w$ & $\alpha$=10,$\beta$=1 & only $\mathcal{L}_r$ & Reference Motion \\
\midrule
$R=4$ & 32.20 & 32.28 & \textbf{32.96} & \textbf{33.01} \\
$R=8$ & 31.10 & 31.12 & \textbf{31.83} & \textbf{32.29} \\
$R=12$ & N.A. & 29.47 & \textbf{29.99} & \textbf{31.23} \\
\bottomrule
\end{tabular}
\end{table}

\begin{table*}[bt!]
\def\doubleunderline#1{\underline{\underline{#1}}}
\centering
\caption{Quantitative comparison of the proposed framework, PICS ($\ell_1$ wavelet)~\cite{lustig2007sparse}, L+S~\cite{LplusS}, 3D MoDL extended from~\cite{MoDL}, CTF-Net~\cite{CTFNet}, GRAFT-Recon~\cite{Hammernik2021} and Unrolled-MCMR~\cite{Pan2022-1} during inference for $R=8, 12, 16$ and $R=20$. Peak Signal to Noise Ratio (PSNR $\downarrow$), Structure Similarity (SSIM $\uparrow$) and Learned Perceptual Image Patch Similarity (LPIPS $\downarrow$) are adopted as evaluation metrics and their averaged value with corresponding standard deviations across all test samples are shown. Methods' average execution time on GPU of reconstructing the sample with spatial resolution $192\times 156$ is logged. PICS ($\ell_1$ wavelet) is carried out on the CPU whose time is not logged. The best results are marked in bold.}
\label{t4}
\begin{tabular}{ccccccccc} 
\toprule
& Metrics & PICS ($\ell_1$ wavelet) & L+S & 3D MoDL & CTF-Net & GRAFT-Recon & Unrolled-MCMR & Proposed\\ \midrule
\multirow{3}*{$R=8$}
& PSNR & 25.264 $\pm$ 2.30 & 26.40 $\pm$ 2.11 & 29.58 $\pm$ 2.88 & 27.41 $\pm$ 2.80 & 29.02 $\pm$ 3.08 & 30.46 $\pm$ 6.23 & \textbf{31.60 $\pm$ 5.41} \\
& SSIM & 0.76 $\pm$ 0.04 & 0.84 $\pm$ 0.04 & 0.91 $\pm$ 0.04 & 0.88 $\pm$ 0.05 & 0.89 $\pm$ 0.05 & 0.90 $\pm$ 0.05 & \textbf{0.92 $\pm$ 0.04} \\
& LPIPS & 0.09 $\pm$ 0.04 & 0.06 $\pm$ 0.02 & 0.02 $\pm$ 0.01 & 0.04 $\pm$ 0.02 & 0.03 $\pm$ 0.02 & 0.03 $\pm$ 0.02 & \textbf{0.02 $\pm$ 0.01} \\
\midrule
\multirow{3}*{$R=12$} 
& PSNR & 19.57 $\pm$ 2.68 & 20.87 $\pm$ 2.65 & 25.37 $\pm$ 2.51 & 25.27 $\pm$ 2.64 & 25.19 $\pm$ 2.53 & 29.41 $\pm$ 4.76 & \textbf{30.16 $\pm$ 4.45} \\
& SSIM & 0.59 $\pm$ 0.09 & 0.70 $\pm$ 0.07 & 0.84 $\pm$ 0.05 & 0.84 $\pm$ 0.05 & 0.80 $\pm$ 0.07 & 0.88 $\pm$ 0.06 & \textbf{0.89 $\pm$ 0.05} \\
& LPIPS & 0.13 $\pm$ 0.04 & 0.09 $\pm$ 0.04 & 0.04 $\pm$ 0.02 & 0.05 $\pm$ 0.03 & 0.07 $\pm$ 0.05 & 0.03 $\pm$ 0.02 & \textbf{0.02 $\pm$ 0.01} \\
\midrule
\multirow{3}*{$R=16$}
& PSNR & 15.94 $\pm$ 2.81 & 16.81 $\pm$ 2.96 & 21.58 $\pm$ 2.05 & 23.67 $\pm$ 2.62 & 23.55 $\pm$ 2.44 & 28.44 $\pm$ 4.20 & \textbf{28.93 $\pm$ 4.08} \\
& SSIM & 0.39 $\pm$ 0.12 & 0.50 $\pm$ 0.13 & 0.75 $\pm$ 0.06 & 0.80 $\pm$ 0.06 & 0.73 $\pm$ 0.08 & 0.86 $\pm$ 0.07 & \textbf{0.87 $\pm$ 0.06} \\
& LPIPS & 0.20 $\pm$ 0.06 & 0.16 $\pm$ 0.06 & 0.06 $\pm$ 0.03 & 0.07 $\pm$ 0.03 & 0.11 $\pm$ 0.06 & 0.03 $\pm$ 0.02 & \textbf{0.03 $\pm$ 0.02} \\
\midrule
\multirow{3}*{$R=20$}
& PSNR & 14.79 $\pm$ 2.70 & 15.27 $\pm$ 2.77 & 19.71 $\pm$ 1.93 & 23.37 $\pm$ 2.68 & 22.66 $\pm$ 2.52 & 27.71 $\pm$ 3.95 & \textbf{27.95 $\pm$ 3.89} \\
& SSIM & 0.31 $\pm$ 0.12 & 0.39 $\pm$ 0.13 & 0.67 $\pm$ 0.07 & 0.78 $\pm$ 0.07 & 0.68 $\pm$ 0.09 & 0.84 $\pm$ 0.08 & \textbf{0.85 $\pm$ 0.07} \\
& LPIPS & 0.23 $\pm$ 0.06 & 0.20 $\pm$ 0.06 & 0.09 $\pm$ 0.05 & 0.08 $\pm$ 0.04 & 0.13 $\pm$ 0.08 & 0.05 $\pm$ 0.03 & \textbf{0.04 $\pm$ 0.03} \\
\midrule
\multirow{1}*{Avg. Time (s)}
& & N.A. & \textbf{1.73 $\pm$ 0.01} & 3.37 $\pm$ 0.01 & 3.00 $\pm$ 0.01 & 5.37 $\pm$ 0.01 & 14.89 $\pm$ 0.03 & 5.49 $\pm$ 0.01 \\
\bottomrule
\end{tabular}
\label{tab:QuantComp}
\end{table*}

\subsection{Baseline comparisons}
\paragraph{\textbf{Application details}} We compare our method to six baseline methods (see~\ref{sub_E}). The deep learning approaches included the proposed network, 3D MoDL~\cite{MoDL} and CTF-Net~\cite{CTFNet} are trained with mixed $R$ training procedure (see~\ref{sub_A}). These methods can achieve their best performance by using this training procedure. The mixed $R$ training procedure is not applied to GRAFT-Recon~\cite{Hammernik2021} and Unrolled-MCMR~\cite{Pan2022-1} since including highly undersampled data gave rise to unstable training and poor reconstruction performance. In practice we found that GRAFT-Recon can achieve the best inference results when using the fixed $R=8$ training compared to all other fixed $R$ trainings. For Unrolled-MCMR, the fixed $R=12$ training is the best training strategy. Thus, for GRAFT-Recon a fixed $R=8$ training is conducted, while for Unrolled-MCMR training we only use $R=12$ data. Besides, we also set their temporal neighborhood to $K=9$ with $\pm 4$ neighboring frames, whilst in their original work they employed all temporal frames which can cause higher warping errors. In order to carry out a fair comparison, the initialization step (refer to \ref{CGSENSE}) is applied for all deep learning-based methods including 3D MoDL, CTF-Net, GRAFT-Recon and Unrolled-MCMR. After training, all six methods are tested on an arbitrary undersampling rate in the retrospective study or the OCMR prospective study with $R=9$.

\begin{figure*}[t!]
	\centering
	\includegraphics[width=1\textwidth]{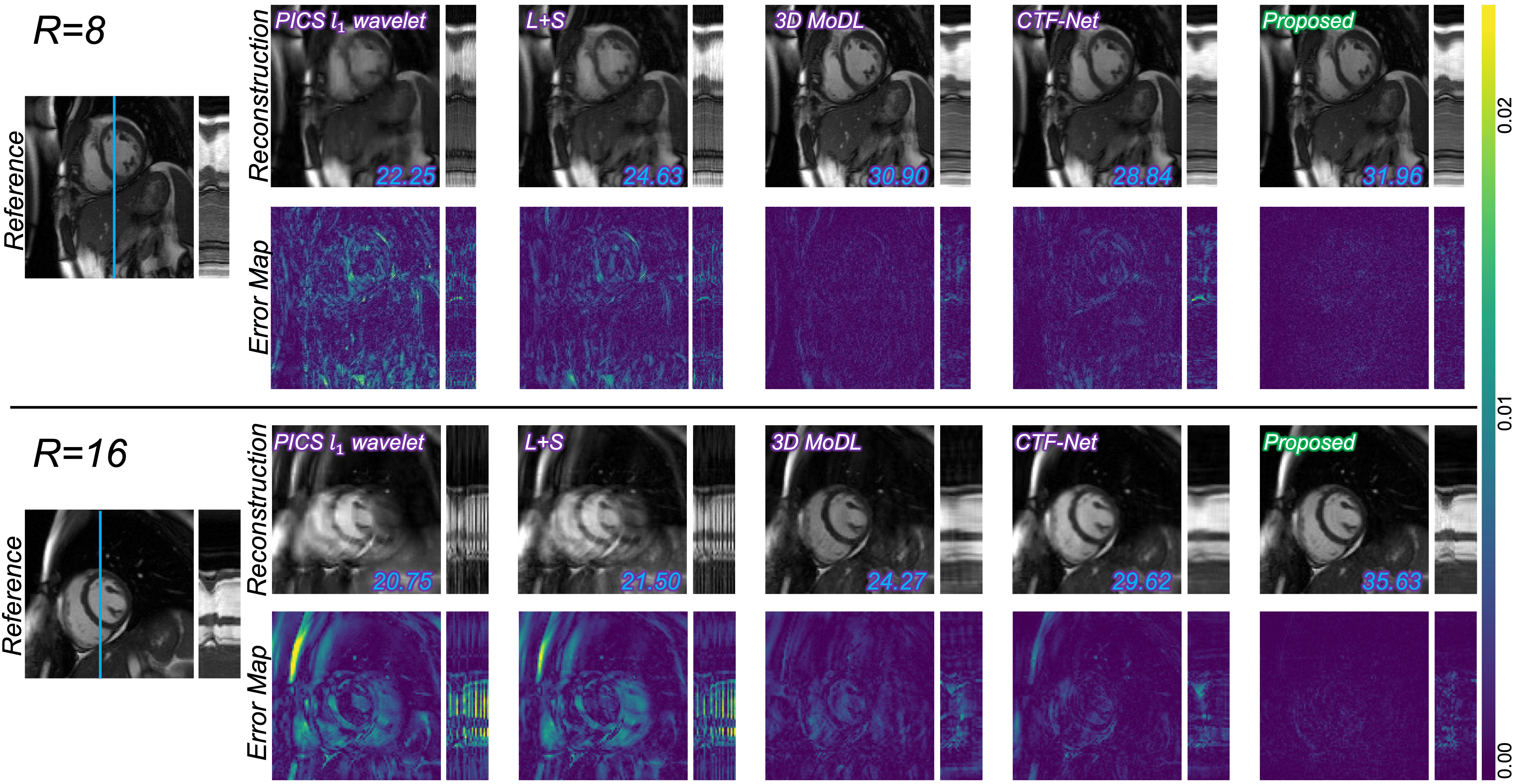}
	\caption{Qualitative comparison of the proposed method to non-MCMR methods including PICS ($\ell_1$ wavelet)~\cite{lustig2007sparse}, L+S~\cite{LplusS}, 3D MoDL extended from~\cite{MoDL} and CTF-Net~\cite{CTFNet} in the $R=8$ (patient with myocarditis) and $R=16$ (healthy subject) accelerated acquisition. The respective PSNR values of the heart region are depicted in the image. Reference images, reconstructed images and their corresponding error maps are demonstrated. The spatial ($x-y$) images are depicted next to the temporal traces ($y-t$) through the middle of the left ventricle. The selected y-axis is marked with a blue line in the reference image.}
	\label{fig3}
 % \vspace{-0.6cm}
\end{figure*}

\paragraph{\textbf{Quantitative analysis}} The quantitative performance of all methods evaluated by metrics PSNR, SSIM, and LPIPS is shown in Table~\ref{t4}. The superior and consistent performance of the proposed method is shown across every single undersampling rate compared to all other baseline methods and regardless of the evaluation metric. It can be noted that learning-based methods e.g. GRAFT-Recon, Unrolled-MCMR, CTF-Net and 3D MoDL outperform conventional methods like PICS ($\ell_1$ wavelet) and L+S. Moreover, Unrolled-MCMR consistently demonstrates the second-best performance because of its usage of reference images in the motion-warping loss function and its unrolled iterative optimization mechanism. The advantage of iterative optimization becomes more prominent for higher acceleration rates. Furthermore, Table~\ref{t5} focuses on the comparative quantitative performance of our method against the best non-MCMR (CTF-Net) and MCMR (Unrolled-MCMR) baselines, specifically for ED and ES frames. These results highlight our method's superior ability to reconstruct ED and ES frames across all acceleration rates, underscoring its potential for practical applications such as indicator extraction.

\paragraph{\textbf{Retrospective qualitative analysis}} The qualitative comparison of two test subjects (healthy subject and patient) between the proposed network and the non-MCMR methods is illustrated in Fig.~\ref{fig3} for undersampling rates of $R=8$ and $R=16$. The corresponding error maps are displayed as well. The proposed network presents a consistent performance in both subjects with the highest PSNR score and lowest residual error. Temporal traces are in good agreement with the fully-sampled reference and cardiac dynamics were recovered by the proposed network. Clinic useful features like papillary muscles are restored clearly without blurring in both cases.

\begin{figure*}[t!]
	\centering
	\includegraphics[width=1\textwidth]{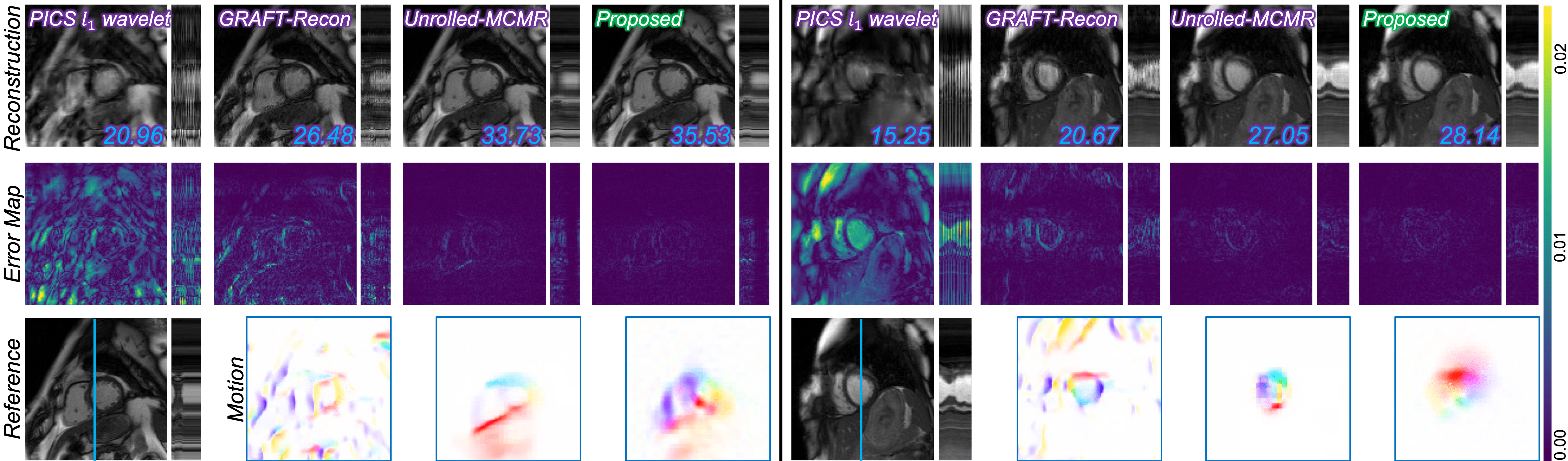}
	\caption{Qualitative comparison of the proposed network to PICS ($\ell_1$ wavelet)~\cite{lustig2007sparse}, GRAFT-Recon~\cite{Hammernik2021} and Unrolled-MCMR~\cite{Pan2022-1} in the $R=12$ (left side, healthy subject) and $R=20$ (right side, patient with myocarditis) accelerated acquisition. The respective PSNR values of the heart region are shown in the image. Reference images, reconstructed images, corresponding error maps and color-wheel-encoded~\cite{colorwheel} motion field visualization are shown. The spatial ($x-y$) images are depicted next to the temporal traces ($y-t$) through the left ventricle. The selected y-axis is marked with a blue line in the reference image.}
	\label{fig4}
 % \vspace{-0.6cm}
\end{figure*}

Further retrospective qualitative comparison of two test subjects (healthy subject and patient) between the proposed framework and other MCMR methods (GRAFT-Recon and Unrolled-MCMR) are demonstrated in Fig. \ref{fig4}. The proposed network outperforms the two compared MCMR methods in both $R=12$ and $R=20$. While the proposed framework is only trained with reconstruction loss without any smoothness terms, it predicts a more meaningful and dense motion field even for $R=20$. The motion estimation from Unrolled-MCMR is sparse and non-smooth, in spite of the usage of smoothness terms during training. The GRAFT-Recon reveals inferior reconstruction due to the motion estimation being artifact-affected resulting in error propagation amongst frames, while the proposed method yields a reconstruction image without any aliasing in both cardiac region and background.

\begin{figure*}[t!]
	\centering
	\includegraphics[width=1\textwidth]{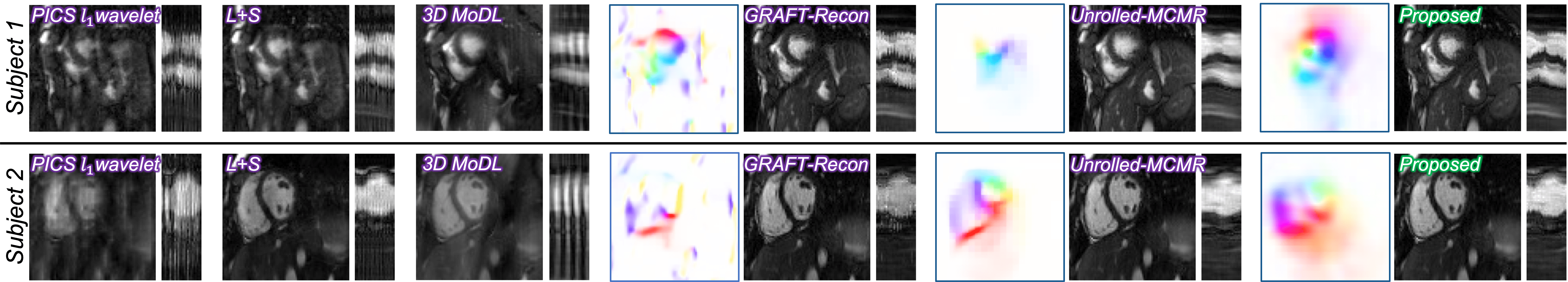}
	\caption{Qualitative comparison of the proposed network to PICS ($\ell_1$ wavelet)~\cite{lustig2007sparse}, L+S~\cite{LplusS}, MoDL~\cite{MoDL}, GRAFT-Recon~\cite{Hammernik2021} and Unrolled-MCMR~\cite{Pan2022-1} in two OCMR prospectively undersampled subjects with the acceleration rate of $R=9$. Reconstructed images in $x-y$ and $y-t$ planes and motion field visualization of the MCMR approaches are illustrated.}
	\label{fig_pros}
 % \vspace{-0.6cm}
\end{figure*}

\paragraph{\textbf{Prospective qualitative analysis}} We further evaluate the proposed method along with the baselines on the real-time prospectively undersampled OCMR data. It is to note that domain shifts occur since we apply the direct inference on the OCMR data without any fine-tuning or transfer learning. The visual comparison based on two subjects is illustrated in Fig. \ref{fig_pros}. CTF-Net cannot be carried out in this experiment due to a memory limitation (data size is increased by more than 4 times compared to the in-house acquired data). Because the spatial resolution is lower than those of the data in the retrospective study, the image quality of the prospective study is not as high as that in the in-house acquired data. It can be observed that the proposed method outperforms the baseline methods with less spatial blurring and streaking artifacts. Relevant diagnostic features such as papillary muscles (as shown in subject 2) are restored without blurring. It is also to note that the generated motion fields of the proposed method are also more dense, smooth and meaningful than the motion fields from other MCMR methods. We can conclude that the proposed method demonstrates generalizability and robust behavior toward subject domain shifts.

% \begin{table}
% \centering
% \caption{The quantitative comparison of the proposed method compared to CTF-Net and Unrolled-MCMR only at \textbf{end-diastolic and end-systolic} frames. The averaged PSNR of all test samples and the standard deviation of the final reconstruction at acceleration rates of 8, 12, 16 and 20 are shown. The best results are marked in bold.}
% \label{t5}
% \begin{tabular}{lccc}
% \toprule
%   & CTF-Net & Unrolled-MCMR & Proposed \\
% \midrule
% $R=8$ & 27.57 $\pm$ 2.98  & 31.80 $\pm$ 7.95 & \textbf{32.77 $\pm$ 6.56} \\
% $R=12$ & 25.25 $\pm$ 2.76 & 30.62 $\pm$ 5.58 & \textbf{31.12 $\pm$ 5.18} \\
% $R=16$ & 23.51 $\pm$ 2.69 & 29.34 $\pm$ 4.12 & \textbf{29.64 $\pm$ 3.99} \\
% $R=20$ & 23.23 $\pm$ 2.80 & 28.33 $\pm$ 3.76 & \textbf{28.45 $\pm$ 3.71} \\
% \bottomrule
% \end{tabular}
% \end{table}

\begin{table}
\centering
\caption{The quantitative comparison of the proposed method compared to CTF-Net and Unrolled-MCMR only at \textbf{end-systolic (ES)} and \textbf{end-diastolic (ED)} frames of the test subjects. The averaged PSNR and standard deviation of the final reconstruction at acceleration rates of 8, 12, 16 and 20 are shown. The best results are marked in bold.}
\label{t5}
\begin{tabular}{lcccc}
\toprule
& & CTF-Net & Unrolled-MCMR & Proposed \\
\midrule
\multirow{2}*{$R=8$}
& ES & 25.63 $\pm$ 2.29 & 28.19 $\pm$ 2.31  & \textbf{30.14 $\pm$ 2.63} \\
& ED & 29.34 $\pm$ 2.29 & 32.24 $\pm$ 2.51 & \textbf{32.88 $\pm$ 2.28} \\
\midrule
\multirow{2}*{$R=12$}
& ES & 23.84 $\pm$ 2.04 & 27.77 $\pm$ 2.32  & \textbf{28.68 $\pm$ 2.43} \\
& ED & 26.45 $\pm$ 2.88 & 31.72 $\pm$ 2.54 & \textbf{31.91 $\pm$ 2.63} \\
\midrule
\multirow{2}*{$R=16$}
& ES & 22.03 $\pm$ 1.82 & 26.97 $\pm$ 2.20  & \textbf{27.66 $\pm$ 2.27} \\
& ED & 24.41 $\pm$ 2.76 & 30.92 $\pm$ 2.40 & \textbf{30.93 $\pm$ 2.60} \\
\midrule
\multirow{2}*{$R=20$}
& ES & 21.89 $\pm$ 2.04 & 25.94 $\pm$ 2.16  & \textbf{26.21 $\pm$ 2.19} \\
& ED & 24.30 $\pm$ 3.06 & 30.10 $\pm$ 2.42 & \textbf{30.10 $\pm$ 2.50} \\
\bottomrule
\end{tabular}
\end{table}

\section{Discussion}\color{black}
MCMR is a powerful and straightforward concept that has been demonstrated for the reconstruction of cardiac CINE~\cite{Cruz2021,CS_M2,Qi2022,Pan2022-1}. However, a wide range of MCMR implementations for CINE is precluded by two major unsolved challenges: high-speed MCMR processing and precise artifact-suppressed cardiac motion estimation. In this work, we proposed a learning-based MCMR framework for CINE imaging that copes with these two problems at once. The fast MCMR is achieved by leveraging the trained network to accelerate the estimation process in inference time, whilst the artifact-suppressed motion estimation is achieved using reconstruction-driven motion estimation. We treat the two sub-tasks as a single entity, in which the training loss is back-propagated end-to-end from the final reconstructed images to the motion estimation input.

\paragraph{\textbf{Training strategy and the role of motion estimation in MCMR}} We performed an ablation study in which the use of an intermediate warping similarity loss is compared to a final reconstruction loss. Results indicate that if the optimization is driven by the final reconstruction loss, not only the reconstruction performance is enhanced, but also the motion prediction is ameliorated. Moreover, using the proposed training strategy can yield closer performance to the "ideal" setting which uses motion predicted from fully-sampled images for the MCMR. 

The motion estimation plays an essential role in MCMR and its quality can directly influence the quality of the final reconstruction. As observed from Fig.~\ref{figFullyUS}, using realistic, dense and detailed motion field (reference motion) estimated from the fully-sampled images can achieve the best reconstruction with the least remaining errors. Using the proposed reconstruction-driven loss $\mathcal{L}_r$ can also achieve a dense and smooth motion field, resembling the reference motion and therefore deliver a closer reconstruction performance to the "ideal" setting. In contrast, using the canonical motion warping loss $\mathcal{L}_w$ results in an underestimated cardiac motion field, and a suboptimal reconstruction with the most residual errors subsequently.

It is also to note that our framework is not limited to certain components such as GRAFT (for motion estimation) and CG-SENSE (for reconstruction) applied in this work. The proposed framework allows for the integration of alternative motion estimation networks and reconstruction methods, in which the back-propagation is carried out through the whole pipeline. This transferability enhances the potential impact of our work, as it can be tailored to a wide range of scenarios in the field.

\paragraph{\textbf{Temporal redundancy and the number of neighboring frames for MCMR}} 
In CMR reconstruction, it is beneficial to consider more frames of the cardiac cycle to conduct the reconstruction because of the adequate temporal redundancy of CMR. In this work, we find the importance of leveraging temporal redundancy increases with the elevation of acceleration rates, and it is less critical to leverage redundancy at lower acceleration rates ($R\leq 8$). This conclusion can be drawn from our experiments on the optimal number of neighboring frames used for MCMR (refer to Table~\ref{t2} and Fig.~\ref{fig2.1}). While at $R=8$ leveraging $\pm2$ temporal frames already yields superior results, the proposed method demands $\pm6$ frames at R=20 to get satisfactory reconstruction. 

Further, there is also a trade-off between the static reconstruction error which is incurred by the lack of redundant information, and the dynamic reconstruction error around the heart which is caused by residual warping errors from neighboring frames. Using $K=9$ is the optimal choice in our case, which can achieve superior results and satisfactory temporal resolution even without the regularization term (refer to Fig.~\ref{fig_lambda_k}). Further, from Fig.~\ref{fig_lambda_k} we can find that the choice of a large $K$ is more critical at the ES frame than at the ED frame. The cardiac motion is larger around ES frame while through-plane motion can also occur more frequently, resulting in more residual registration and further reconstruction errors with averaging effect. This conclusion can also be drawn from Table~\ref{t5} since the reconstruction PSNR at ES is inferior to ED. This indicates the importance of using a proper amount of neighboring frames $K$ for the reconstruction.
 It is important to note that this phenomenon is not only occurring in our proposed approach but is generic for any MCMR method. Based on these results, we set a fixed number of neighboring frames in this work. In the future, we can use soft temporal weighting window (e.g. gaussian kernel) with learnable bandwidth so that our approach can self-adapt to the optimal value for different application scenarios. 

 \paragraph{\textbf{Relevance to free-breathing scan}} Besides the superior performance at low acceleration rates ($R=8,12$), the proposed approach also presents consistent results in highly undersampled cases ($R=16,20$) over baseline methods and markedly outperforms non-MCMR methods. This enhancement is particularly relevant in free-breathing scans. Typically, in these scans, not only is the cardiac phase binned, but k-space also needs to be binned to account for respiratory motion during data collection. Given the same data acquisition duration, fewer k-space lines can be collected in the free-breathing scan for every cardiac phase than in the breath-hold scan due to the extra respiratory motion binning. Therefore, the k-space data is further undersampled. The evidence shown in Table~\ref{tab:QuantComp}, Fig.~\ref{fig3} and \ref{fig4} demonstrates the reliable performance of our method in highly undersampled data, highlighting our method's significant potential for further implementation in free-breathing scan settings.

\paragraph{\textbf{MCMR comparison}} The superior results of the proposed method against Unrolled-MCMR can be attributed to two major reasons. First, we carry out just a single but more effective optimization instead of applying alternating updates of motion fields and image reconstructions. It should be noted that our proposed method can also be extended as an iterative unrolled optimization but at the cost of prolonged training and test time. Second, the proposed method can carry out artifact-suppressed motion estimation (refer to the motion fields in Fig. \ref{fig4}). The proposed \textit{Motion-Compensated Reconstruction Block} can be regarded as a transformation operator which extends the motion estimation procedure from image space to k-space. Although Eq. \eqref{eq12} presents a loss function that forces the framework to generate a reference resembled reconstruction, it can also be interpreted as a warping loss function which warps a set of undersampled images by the estimated motion to the target images while ensuring consistency to acquired k-space samples.

\paragraph{\textbf{CMR motion estimation/registration}} Our proposed approach provides another perspective on solving the cardiac motion estimation/registration problem. Cardiac motion estimation/registration can not only be used inside the MCMR framework for reconstruction but can also be applied for cardiac feature tracking to evaluate myocardial strain and functional analysis~\cite{FeatureTrack1,FeatureTrack2} or to facilitate cardiac segmentation tasks~\cite{Qin2018JointLO}. Our proposed method can be recast to a motion estimation/registration method with two major benefits compared to the conventional motion estimation/registration methods. First, we only need a single loss term (Eq. \eqref{eq12}) to generate smooth and realistic motion fields without introducing regularizers on motion. Second, we can predict high-quality cardiac motion directly from highly undersampled MR data. It is also conceivable that we do not need visually appealing MR images for the extraction and quantifícation of clinical parameters (e.g. left ventricular function). A potential synergistic approach for jointly reconstructing, analyzing (e.g. segmentation or motion tracking) and interpreting the cardiac CINE imaging will be developed further based on this study.  

\paragraph{\textbf{Transfer to self-supervised learning setting}} 
Currently, the introduced method is established on the supervised learning setting which necessitates fully-sampled reference data to guide the training. The recent advancements in self-supervised learning (SSL) in MR reconstruction enable the training when the fully-sampled images are absent~\cite{yaman2020self,hu2021self,yaman2022zeroshot,acar2021self}. These methods proposed a training strategy to split the k-space of the present undersampled data into two non-overlapped segments, one is used as the reference for training while the other is to enforce data consistency. This training strategy can be directly applied to the dynamic MR reconstruction~\cite{acar2021self} and also to ours. However, the trade-off problem is to be considered since one has to tolerate performance degradation while transferring the framework into the SSL setting as pointed out in these aforementioned works.

\paragraph{\textbf{Limitation}}  we also acknowledge some limitations of our work. First, the motion estimation is based on the backbone of GRAFT. It conducts $N \times K$ computations to reconstruct the cardiac cycle with $N$ frames which are suboptimal regarding estimation speed and memory usage. In future work, we will attempt to build a more efficient and lightweight group-wise motion estimator to accelerate the reconstruction process further. Second, in this work we have not investigated the impact of introducing preconditioning~\cite{Pruessmann2001} into CG-SENSE, which is assigned to our future work. Moreover, the clinical utility of the proposed method has only been justified in qualitative evaluation (refer to well-reconstructed diagnostic features e.g. papillary muscles in retrospective and prospective studies in Fig.~\ref{fig3} and Fig.~\ref{fig_pros}) and in quantitative ED and ES frames image error evaluation. A study using robust segmentation networks e.g.~\cite{chen2020improving,chen2020realistic} to provide a direct indication of tolerable accelerations for standard functional assessments will be evaluated in future work. Finally, currently our method is applied in 2D CMR and we have not compared our proposed method with 3D commercial solutions e.g. Philips compressed-sense reconstruction. In the future, we will investigate the transition to 3D CMR and provide more concrete evidence of the applicability and effectiveness of our methodology compared to commercial solutions.

\section{Conclusion}

In this work, we proposed a learning-based MCMR framework for CINE imaging. We introduce a mechanism that solves the MCMR problem as a single entity and drives the motion estimation directly from the final reconstruction perspective. The training loss is back-propagated through the whole pipeline and the framework is optimized end-to-end without breaking into two sub-tasks and without using any regularization/smoothness loss terms. We find out that using a smaller neighboring frames number to conduct MCMR can achieve better results than using all sequence frames. Our method shows consistent and robust performance throughout all conducted experiments and outperforms all baseline methods. We have confidence that the developed method for cardiac CINE imaging can also be generalized and applied to other reconstruction applications.
% In this work, we proposed an end-to-end MCMR framework for CINE imaging with a straightforward and efficient training strategy. Instead of using conventional image warping similarity as the intermediate loss function to predict the motion, we introduce a mechanism that regards the MCMR problem as a single entity and drives the motion estimation directly from the final reconstruction perspective. Our method shows consistent performance throughout all conducted experiments and outperforms all baseline methods. We have confidence that the developed method for cardiac CINE imaging can also be generalized and applied to other reconstruction applications.

\appendices

% \begin{thebibliography}{00}
\bibliographystyle{unsrt}

\bibliography{bibliography_shortened.bib}

\begin{thebibliography}{10}

\bibitem{TSENSE}
P.~Kellman, F.~H. Epstein, and E.~R. McVeigh.
\newblock Adaptive sensitivity encoding incorporating temporal filtering
  (tsense)†.
\newblock {\em Magnetic Resonance in Medicine}, 45(5):846--852, 2001.

\bibitem{TGRAPPA}
F.~A. Breuer, P.~Kellman, M.~A. Griswold, and P.~M. Jakob.
\newblock Dynamic autocalibrated parallel imaging using temporal grappa
  (tgrappa).
\newblock {\em Magnetic Resonance in Medicine}, 53(4):981--985, 2005.

\bibitem{k-t-focuss}
H.~Jung, K.~Sung, K.~S. Nayak, E.~Y. Kim, and Ye~J. C.
\newblock k-t focuss: A general compressed sensing framework for high
  resolution dynamic mri.
\newblock {\em Magnetic Resonance in Medicine}, 61(1):103--116, 2009.

\bibitem{CompCSPI}
R.~Otazo, D.~Kim, L.~Axel, and D.~K. Sodickson.
\newblock Combination of compressed sensing and parallel imaging for highly
  accelerated first-pass cardiac perfusion mri.
\newblock {\em Magnetic Resonance in Medicine}, 64(3):767--776, 2010.

\bibitem{Batchelor2005}
P.~G. Batchelor, D.~Atkinson, P.~Irarrazaval, D.~L.~G Hill, Hajnal J., and
  Larkman D.
\newblock Matrix description of general motion correction applied to multishot
  images.
\newblock {\em Magnetic Resonance in Medicine}, 54:1273–1280, 2005.

\bibitem{LplusS}
R.~Otazo, E.~Candès, and D.~K. Sodickson.
\newblock Low-rank plus sparse matrix decomposition for accelerated dynamic mri
  with separation of background and dynamic components.
\newblock {\em Magnetic Resonance in Medicine}, 73(3):1125--1136, 2015.

\bibitem{Huang21}
W.~Huang and et~al.
\newblock Deep low-rank plus sparse network for dynamic mr imaging.
\newblock {\em Medical Image Analysis}, 73:102190, 2021.

\bibitem{Kuestner2020CINENet}
T.~K{\"{u}}stner, N.~Fuin, K.~Hammernik, and et~al.
\newblock {CINENet: deep learning-based 3D cardiac CINE MRI reconstruction with
  multi-coil complex-valued 4D spatio-temporal convolutions}.
\newblock {\em Scientific Reports}, 10(1):1--13, 2020.

\bibitem{dl-espirit}
C.~M. Sandino, P.~Lai, S.~S. Vasanawala, and J.~Y. Cheng.
\newblock Accelerating cardiac cine mri using a deep learning-based espirit
  reconstruction.
\newblock {\em Magnetic Resonance in Medicine}, 85(1), 2021.

\bibitem{kt_slr}
S.~G. Lingala, Y.~Hu, E.~DiBella, and M.~Jacob.
\newblock Accelerated dynamic mri exploiting sparsity and low-rank structure:
  k-t slr.
\newblock {\em IEEE Transactions on Medical Imaging}, 30(5):1042--1054, 2011.

\bibitem{Qin2019}
C.~Qin and et~al.
\newblock k-t next: Dynamic mr image reconstruction exploiting spatio-temporal
  correlations.
\newblock In {\em Medical Image Computing and Computer Assisted Intervention},
  pages 505--513, 2019.

\bibitem{CTFNet}
C.~Qin and et~al.
\newblock Complementary time-frequency domain networks for dynamic parallel mr
  image reconstruction.
\newblock {\em Magnetic Resonance in Medicine}, 86(6):3274--3291, 2021.

\bibitem{ghodrati2021retrospective}
V.~Ghodrati, M.~Bydder, F.~Ali, C.~Gao, A.~Prosper, K.~Nguyen, and P.~Hu.
\newblock Retrospective respiratory motion correction in cardiac cine mri
  reconstruction using adversarial autoencoder and unsupervised learning.
\newblock {\em NMR in Biomedicine}, 34(2):e4433, 2021.

\bibitem{kustner2019retrospective}
T.~K{\"u}stner, K.~Armanious, J.~Yang, B.~Yang, F.~Schick, and S.~Gatidis.
\newblock Retrospective correction of motion-affected mr images using deep
  learning frameworks.
\newblock {\em Magnetic resonance in medicine}, 82(4):1527--1540, 2019.

\bibitem{klein2009elastix}
S.~Klein, M.~Staring, K.~Murphy, and J.~Pluim.
\newblock Elastix: a toolbox for intensity-based medical image registration.
\newblock {\em IEEE Transactions on Medical Imaging}, 29(1):196--205, 2009.

\bibitem{Modat2010}
M.~Modat and et~al.
\newblock Fast free-form deformation using graphics processing units.
\newblock {\em Computer Methods and Programs in Biomedicine}, 98(3):278--284,
  2010.

\bibitem{vercauteren2009diffeomorphic}
T.~Vercauteren, X.~Pennec, A.~Perchant, and N.~Ayache.
\newblock Diffeomorphic demons: Efficient non-parametric image registration.
\newblock {\em Neuro Image}, 45(1):S61--S72, 2009.

\bibitem{registration_review}
X.~Chen, A.~Diaz-Pinto, N.~Ravikumar, and A.~F. Frangi.
\newblock Deep learning in medical image registration.
\newblock {\em Progress in Biomedical Engineering}, 3(1):012003, feb 2021.

\bibitem{voxelmorph1}
G.~Balakrishnan, A.~Zhao, M.~R. Sabuncu, J.~Guttag, and A.~Dalca.
\newblock An unsupervised learning model for deformable medical image
  registration.
\newblock In {\em 2018 IEEE/CVF Conference on Computer Vision and Pattern
  Recognition}, pages 9252--9260, 2018.

\bibitem{voxelmorph2}
G.~Balakrishnan, A.~Zhao, M.~R. Sabuncu, J.~Guttag, and A.~Dalca.
\newblock Voxelmorph: A learning framework for deformable medical image
  registration.
\newblock {\em IEEE Transactions on Medical Imaging}, 38(8):1788--1800, 2019.

\bibitem{Morales2019}
M.~A. Morales, D.~Izquierdo-Garcia, I.~Aganj, J.~Kalpathy-Cramer, B.~R. Rosen,
  and C.~Catana.
\newblock Implementation and validation of a three-dimensional cardiac motion
  estimation network.
\newblock In {\em Radiol Artificial Intelligence}, volume~4, 2019.

\bibitem{MPN2020}
H.~Yu, X.~Chen, H.~Shi, T.~Chen, T.~S. Huang, and S.~Sun.
\newblock Motion pyramid networks for accurate and efficient cardiac motion
  estimation.
\newblock {\em CoRR}, abs/2006.15710, 2020.

\bibitem{Pan2021}
J.~Pan, D.~Rueckert, T.~K{\"u}stner, and K.~Hammernik.
\newblock Efficient image registration network for non-rigid cardiac motion
  estimation.
\newblock In {\em Machine Learning for Medical Image Reconstruction}, pages
  14--24, 2021.

\bibitem{Qi2021-2}
H.~Qi and et~al.
\newblock End-to-end deep learning nonrigid motion-corrected reconstruction for
  highly accelerated free-breathing coronary mra.
\newblock {\em Magnetic Resonance in Medicine}, 86(1):1983--1996, 2021.

\bibitem{LAPNet}
T.~Küstner and et~al.
\newblock Lapnet: Non-rigid registration derived in k-space for magnetic
  resonance imaging.
\newblock {\em IEEE Transactions on Medical Imaging}, 40(12):3686--3697, 2021.

\bibitem{Hammernik2021}
K.~Hammernik, J.~Pan, D.~Rueckert, and T.~K{\"u}stner.
\newblock Motion-guided physics-based learning for cardiac mri reconstruction.
\newblock {\em Asilomar Conference on Signals, Systems, and Computers}, 2021.

\bibitem{Qi2022}
J.~Yang, T.~Küstner, P.~Hu, P.~Liò, and H.~Qi.
\newblock End-to-end deep learning of non-rigid groupwise registration and
  reconstruction of dynamic mri.
\newblock {\em Frontiers in Cardiovascular Medicine}, 9, 2022.

\bibitem{Prost1}
G.~Cruz, D.~Atkinson, M.~Henningsson, R.~Botnar, and C.~Prieto.
\newblock Highly efficient nonrigid motion-corrected 3d whole-heart coronary
  vessel wall imaging.
\newblock {\em Magnetic Resonance in Medicine}, 77(5):1894--1908, 2017.

\bibitem{Prost2}
A.~Bustin and et~al.
\newblock 3d whole-heart isotropic sub-millimeter resolution coronary magnetic
  resonance angiography with non-rigid motion-compensated prost.
\newblock {\em Journal of Cardiovascular Magnetic Resonance}, 22(1), 2020.

\bibitem{MUNOZ202210}
C.~Munoz, H.~Qi, G.~Cruz, T.~Küstner, R.~Botnar, and C.~Prieto.
\newblock Self-supervised learning-based diffeomorphic non-rigid motion
  estimation for fast motion-compensated coronary mr angiography.
\newblock {\em Magnetic Resonance Imaging}, 85:10--18, 2022.

\bibitem{zou2022dynamic}
Q.~Zou, L.~Torres, S.~Fain, N.~Higano, A.~Bates, and M.~Jacob.
\newblock Dynamic imaging using motion-compensated smoothness regularization on
  manifolds (moco-storm).
\newblock {\em Physics in medicine \& biology}, 67(14):144001, 2022.

\bibitem{Cruz2021}
G.~Cruz, K.~Hammernik, T.~Kuestner, D.~Rueckert, R.~Botnar, and C.~Prieto.
\newblock One-heartbeat cardiac cine imaging via jointly regularized non-rigid
  motion corrected reconstruction.
\newblock In {\em Proc. International Society for Magnetic Resonance in
  Medicine (ISMRM)}, page 0070, 2021.

\bibitem{Odille2008}
F.~Odille, P.~A. Vuissoz, P.~Y. Marie, and J.~Felblinger.
\newblock Generalized reconstruction by inversion of coupled systems (grics)
  applied to free-breathing mri.
\newblock {\em Magnetic Resonance in Medicine}, 60:146--157, 2008.

\bibitem{Odille2016}
F.~Odille and et~al.
\newblock Joint reconstruction of multiple images and motion in mri:
  Application to free-breathing myocardial ${\rm t}_{2}$ quantification.
\newblock {\em IEEE Transactions on Medical Imaging}, 35(1):197--207, 2016.

\bibitem{NiftyReconJoint}
J.~Royuela-del Val, L.~Cordero-Grande, F.~Simmross-Wattenberg,
  M.~Martín-Fernández, and C.~Alberola-López.
\newblock Nonrigid groupwise registration for motion estimation and
  compensation in compressed sensing reconstruction of breath-hold cardiac cine
  mri.
\newblock {\em Magnetic Resonance in Medicine}, 75(4):1525--1536, 2016.

\bibitem{FlowReconJoint}
N.~Zhao, D.~O’Connor, A.~Basarab, D.~Ruan, and K.~Sheng.
\newblock Motion compensated dynamic mri reconstruction with local affine
  optical flow estimation.
\newblock {\em IEEE Transactions on Biomedical Engineering}, 66(11):3050--3059,
  2019.

\bibitem{CS_M}
A.~I. Aviles-Rivero, N.~Debroux, G.~Williams, M.~J. Graves, and C.~Schönlieb.
\newblock Compressed sensing plus motion (cs + m): A new perspective for
  improving undersampled mr image reconstruction.
\newblock {\em Medical Image Analysis}, 68:101933, 2021.

\bibitem{CS_M2}
T.~Schmoderer, A.~I Aviles-Rivero, V.~Corona, N.~Debroux, and C.~Schönlieb.
\newblock Learning optical flow for fast {MRI} reconstruction.
\newblock {\em Inverse Problems}, 37(9):095007, aug 2021.

\bibitem{Pan2022-1}
J.~Pan, D.~Rueckert, T.~K{\"u}stner, and K.~Hammernik.
\newblock Learning-based and unrolled motion-compensated reconstruction
  for cardiac mr cine imaging.
\newblock In {\em Medical Image Computing and Computer Assisted Intervention},
  pages 686--696, 2022.

\bibitem{LAPRecon}
T.~Küstner and et~al.
\newblock Deep-learning based motion-corrected image reconstruction in 4d
  magnetic resonance imaging of the body trunk.
\newblock In {\em 2020 Asia-Pacific Signal and Information Processing
  Association Annual Summit and Conference (APSIPA ASC)}, pages 976--985, 2020.

\bibitem{chen2020ocmr}
C.~Chen, Y.~Liu, P.~Schniter, M.~Tong, K.~Zareba, O.~Simonetti, L.~Potter, and
  R.~Ahmad.
\newblock Ocmr (v1. 0)--open-access multi-coil k-space dataset for
  cardiovascular magnetic resonance imaging.
\newblock {\em arXiv preprint arXiv:2008.03410}, 2020.

\bibitem{Pruessmann2001}
K.~P. Pruessmann, M.~Weiger, P.~B{\"o}rnert, and P.~Boesiger.
\newblock Advances in sensitivity encoding with arbitrary k-space trajectories.
\newblock {\em Magnetic Resonance in Medicine}, pages 638--651, 2001.

\bibitem{sandino2020compressed}
C.~Sandino, J.~Cheng, F.~Chen, M.~Mardani, J.~Pauly, and S.~Vasanawala.
\newblock Compressed sensing: From research to clinical practice with deep
  neural networks: Shortening scan times for magnetic resonance imaging.
\newblock {\em IEEE signal processing magazine}, 37(1):117--127, 2020.

\bibitem{teed2020raft}
Z.~Teed and J.~Deng.
\newblock Raft: Recurrent all-pairs field transforms for optical flow.
\newblock In {\em European Conference on Computer Vision (ECCV)}, pages
  402--419. Springer, 2020.

\bibitem{MoDL}
H.~K. Aggarwal, M.~P. Mani, and M.~Jacob.
\newblock Model based image reconstruction using deep learned priors (modl).
\newblock In {\em IEEE International Symposium on Biomedical Imaging (ISBI)},
  pages 671--674, 2018.

\bibitem{CG}
W.~H. Press, S.~A. Teukolsky, W.~T. Vetterling, and B.~P. Flannery.
\newblock {\em Numerical Recipes in C}.
\newblock Cambridge University Press, Cambridge, USA, second edition, 1992.

\bibitem{Ahmad2015}
R.~Ahmad, H.~Xue, S.~Giri, Y.~Ding, J.~Craft, and O.~P. Simonetti.
\newblock Variable density incoherent spatiotemporal acquisition (vista) for
  highly accelerated cardiac mri.
\newblock {\em Magnetic Resonance in Medicine}, 74(5):1266--1278, 2015.

\bibitem{chen2019sparsity}
C.~Chen and et~al.
\newblock Sparsity adaptive reconstruction for highly accelerated cardiac mri.
\newblock {\em Magnetic resonance in medicine}, 81(6):3875--3887, 2019.

\bibitem{uecker2014espirit}
M.~Uecker, P.~Lai, M.~Murphy, P.~Virtue, M.~Elad, J.~Pauly, S.~Vasanawala, and
  M.~Lustig.
\newblock Espirit—an eigenvalue approach to autocalibrating parallel mri:
  where sense meets grappa.
\newblock {\em Magnetic resonance in medicine}, 71(3):990--1001, 2014.

\bibitem{adamw}
I.~Loshchilov and F.~Hutter.
\newblock Decoupled weight decay regularization.
\newblock {\em arXiv preprint arXiv:1711.05101}, 2017.

\bibitem{lustig2007sparse}
M.~Lustig, D.~Donoho, and J.~Pauly.
\newblock Sparse mri: The application of compressed sensing for rapid mr
  imaging.
\newblock {\em Magnetic Resonance in Medicine: An Official Journal of the
  International Society for Magnetic Resonance in Medicine}, 58(6):1182--1195,
  2007.

\bibitem{Wang2004}
Z.~Wang, A.~C. Bovik, H.~R. Sheikh, and E.~P. Simoncelli.
\newblock Image quality assessment: from error visibility to structural
  similarity.
\newblock {\em IEEE Transaction of Image Process}, 13(4):600--612, 2004.

\bibitem{LPIPS}
R.~Zhang, P.~Isola, A.~A. Efros, E.~Shechtman, and O.~Wang.
\newblock The unreasonable effectiveness of deep features as a perceptual
  metric.
\newblock In {\em Conference on Computer Vision and Pattern Recognition}, 2018.

\bibitem{ChenCardiacReview}
C.~Chen and et~al.
\newblock Deep learning for cardiac image segmentation: A review.
\newblock {\em Frontiers in Cardiovascular Medicine}, 7, 2020.

\bibitem{colorwheel}
S.~Baker and et~al.
\newblock A database and evaluation methodology for optical flow.
\newblock In {\em 2007 IEEE 11th International Conference on Computer Vision},
  pages 1--8, 2007.

\bibitem{FeatureTrack1}
R.~J. Taylor and et~al.
\newblock {Myocardial strain measurement with feature-tracking cardiovascular
  magnetic resonance: normal values}.
\newblock {\em European Heart Journal - Cardiovascular Imaging},
  16(8):871--881, 02 2015.

\bibitem{FeatureTrack2}
G.~Pedrizzetti, P.~Claus, P.~J. Kilner, and E.~Nagel.
\newblock Principles of cardiovascular magnetic resonance feature tracking and
  echocardiographic speckle tracking for informed clinical use.
\newblock {\em Journal of Cardiovascular Magnetic Resonance}, 18(1):51, Aug
  2016.

\bibitem{Qin2018JointLO}
C.~Qin and et~al.
\newblock Joint learning of motion estimation and segmentation for cardiac mr
  image sequences.
\newblock In {\em Medical Image Computing and Computer Assisted Intervention},
  2018.

\bibitem{yaman2020self}
B.~Yaman, S.~Hosseini, S.~Moeller, J.~Ellermann, K.~U{\u{g}}urbil, and
  M.~Ak{\c{c}}akaya.
\newblock Self-supervised learning of physics-guided reconstruction neural
  networks without fully sampled reference data.
\newblock {\em Magnetic resonance in medicine}, 84(6):3172--3191, 2020.

\bibitem{hu2021self}
C.~Hu, C.~Li, H.~Wang, Q.~Liu, H.~Zheng, and S.~Wang.
\newblock Self-supervised learning for mri reconstruction with a parallel
  network training framework.
\newblock In {\em Medical Image Computing and Computer Assisted Intervention},
  pages 382--391. Springer, 2021.

\bibitem{yaman2022zeroshot}
B.~Yaman, S.~Hosseini, and M.~Akcakaya.
\newblock Zero-shot self-supervised learning for {MRI} reconstruction.
\newblock In {\em International Conference on Learning Representations}, 2022.

\bibitem{acar2021self}
M.~Acar, T.~{\c{C}}ukur, and {\.I}.~{\"O}ks{\"u}z.
\newblock Self-supervised dynamic mri reconstruction.
\newblock In {\em Machine Learning for Medical Image Reconstruction: 4th
  International Workshop, MLMIR 2021, Held in Conjunction with MICCAI 2021},
  pages 35--44. Springer, 2021.

\bibitem{chen2020improving}
C.~Chen and et~al.
\newblock Improving the generalizability of convolutional neural network-based
  segmentation on cmr images.
\newblock {\em Frontiers in cardiovascular medicine}, 7:105, 2020.

\bibitem{chen2020realistic}
C.~Chen and et~al.
\newblock Realistic adversarial data augmentation for mr image segmentation.
\newblock In {\em Medical Image Computing and Computer Assisted Intervention},
  pages 667--677. Springer, 2020.

\end{thebibliography}
% \end{thebibliography}

\end{document}